\pdfoutput=1
\documentclass[11pt,twoside,a4paper,cmspaper,final,collab]{cms-tdr}
\def\svnVersion{b3c86a3}\def\svnDate{2025/04/17}\def\cmsCernNoTag{CERN-EP-2024-325}\def\cmsCernDate{\today}\def\cmsMessage{Published in the Journal of Instrumentation as \href{http://dx.doi.org/10.1088/1748-0221/20/04/P04021}{\doi{10.1088/1748-0221/20/04/P04021}.}}
\begin{document}\cmsNoteHeader{MUO-24-001}

\newcommand{\jpsi}{\PJGy}
\newcommand{\jpsimm}   {\ensuremath{\PJGy\to \Pgm\Pgm}\xspace}
\newcommand{\bsmm}{\ensuremath{\PBzs\to\Pgm\Pgm}\xspace}
\newcommand{\tautrimu}{\ensuremath{\PGt \to 3\PGm}\xspace}
\newcommand{\abseta}{\ensuremath{\abs{\eta}}\xspace}
\newcommand{\Wjets}{\ensuremath{\PW+\text{jets}}\xspace}
\newcommand{\kspipi}    {\ensuremath{\PKzS\to\PGpp\PGpm}\xspace}
\newcommand{\ksmupi}    {\ensuremath{\PKzS\to\Pgm\Pgn\PGp}\xspace}
\newcommand{\ks}        {\ensuremath{\PKzS}\xspace}
\newcommand{\phikk}     {\ensuremath{\PGf\to \PK\PK}\xspace}
\newcommand{\phimuk}    {\ensuremath{\PGf\to\Pgm\Pgn \PK}\xspace}
\newcommand{\pimunu}{\ensuremath{\PGp \to \Pgm\PGn}\xspace}
\newcommand{\kmunu}{\ensuremath{\PK \to \Pgm\PGn}\xspace}
\providecommand{\cmsTable}[1]{\resizebox{\textwidth}{!}{#1}}

\cmsNoteHeader{MUO-24-001}

\title{Identification of low-momentum muons in the CMS detector using multivariate techniques in proton-proton collisions at \texorpdfstring{$\sqrt{s} = 13.6\TeV$}{sqrt(s) = 13.6 TeV}}

\date{\today}

\abstract{
``Soft'' muons with a transverse momentum below 10\GeV are featured in many processes studied by the CMS experiment, such as decays of heavy-flavor hadrons or rare tau lepton decays. Maximizing the selection efficiency for these muons, while simultaneously suppressing backgrounds from long-lived light-flavor hadron decays, is therefore important for the success of the CMS physics program. Multivariate techniques have been shown to deliver better muon identification performance than traditional selection techniques. To take full advantage of the large data set currently being collected during Run~3 of the CERN LHC, a new multivariate classifier based on a gradient-boosted decision tree has been developed. It offers a significantly improved separation of signal and background muons compared to a similar classifier used for the analysis of the Run~2 data. The performance of the new classifier is evaluated on a data set collected with the CMS detector in 2022 and 2023, corresponding to an integrated luminosity of 62\fbinv.
}
\hypersetup{
pdfauthor={CMS Collaboration},
pdftitle={Identification of low-momentum muons in the CMS detector using multivariate techniques in proton-proton collisions at sqrt(s) = 13.6 TeV},
pdfsubject={CMS},
pdfkeywords={CMS, muon identification, machine learning, multivariate, Run~3}}

\maketitle 

\section{Introduction}
\label{sec:intro}
The CMS Collaboration pursues a broad and comprehensive physics program, exploiting the proton-proton ($\Pp\Pp$) collisions delivered by the CERN LHC to probe particle interactions across energy scales ranging from a few {\GeVns} to several {\TeVns}. A key part of this program is the study of interactions resulting in the production of ``soft'' muons with a transverse momentum $2<\pt < 10\GeV$. Examples of this are the production and decay of heavy-flavor hadrons, such as \bsmm~\cite{CMS:2019bbr,CMS:2022mgd}, or searches for decays that are rare or prohibited in the standard model of particle physics but might be enhanced by extensions to it, \eg, \tautrimu~\cite{CMS:2023iqy, Marciano:2008zz,Raidal:2008jk,Arganda:2005ji}. 

For the success of these studies, it is crucial to select muons produced in these interactions with high efficiency, while effectively rejecting background events. For soft muons, the dominant background arises from the decay in flight of long-lived charged pions and kaons to muons, \ie, \pimunu and \kmunu decays. As a result of the small mass difference between these hadrons and the muon, as well as their significant Lorentz boost, such decays tend to have very small deflections in the trajectory at the decay point, making it difficult to distinguish them from signal muons.

Multivariate (MVA) techniques often outperform traditional approaches based on sets of individual selection criteria~\cite{CMS:2012nsv,CMS:2018rym} (``cut-based identification (ID)''). Such an MVA approach has already been applied to this classification problem in the analysis of the data collected by CMS during Run~2 of the LHC from 2016 to 2018. The resulting ``Run~2 soft-muon MVA'' is a boosted decision tree (BDT) trained initially to be used with the data collected in 2016 using the \textsc{tmva} package~\cite{Hocker:2007ht}, focusing on muons from $\bsmm$ decays with $\pt > 4\GeV$ and an absolute value of pseudorapidity \abseta less than 1.4~\cite{CMS:2019bbr}. It has been widely used in the identification of soft muons outside this restricted phase space in CMS analyses, where it does not perform as optimally, as illustrated by the performance measurements shown below.

Since 2016, there have been significant changes in the CMS detector configuration~\cite{CMS:2008xjf, CMS:2023gfb} and the LHC beam conditions. Among these, the most notable change was the installation of an upgraded pixel subdetector in 2017~\cite{Phase1Pixel} and the increase in the average number of simultaneous $\Pp\Pp$ collisions per bunch crossing (pileup), which rose from 27 in 2016 to 46 in 2022 and 52 in 2023. To address these new conditions, an MVA classifier for soft muon identification in Run~3, which has been ongoing since 2022 to the present, known as the ``Run~3 soft-muon MVA'', has been developed and optimized. Unlike the Run~2 soft-muon MVA from 2016, the Run~3 version was also designed to improve the acceptance for muons with \pt in the range 2--4\GeV and for higher \abseta up to 2.4. This will increase the sensitivity of analyses targeting the physics processes outlined above, allowing us to better utilize the large data set collected during Run~3.

The performance of the Run~3 soft-muon MVA, in terms of selection efficiency and suppression of \PGp and \PK meson decays, is validated using data collected by CMS in 2022 and 2023, corresponding to an integrated luminosity of 62\fbinv~\cite{CMS:LUM-22-001}.

This paper is structured as follows. The CMS detector is described in Section~\ref{sec:cms}, followed by a description of the muon reconstruction in Section~\ref{sec:muons}. The data used for training and performance validation are described in Section~\ref{sec:samples}. An overview of the model design and training is given in Section~\ref{sec:models} and the performance validation using collision data is described in Section~\ref{sec:performance}. A summary of the performance improvements of the new MVA technique is given in Section~\ref{sec:summary}.

\section{The CMS detector}
\label{sec:cms}
The central feature of the CMS apparatus is a superconducting solenoid of 6\unit{m} internal diameter, providing a magnetic field of 3.8\unit{T}. Within the solenoid volume are a silicon pixel and strip tracker, a lead tungstate crystal electromagnetic calorimeter, and a brass and scintillator hadron calorimeter, each composed of a barrel and two endcap sections. Forward calorimeters extend the pseudorapidity coverage provided by the barrel and endcap detectors. Muons are reconstructed in gas-ionization detectors embedded in the steel flux-return yoke outside the solenoid. 

The silicon tracker measures charged particles within the pseudorapidity range $\abs{\eta} < 3.0$. During the LHC running period when the data used in this paper were recorded, the silicon tracker consisted of 1856 silicon pixel and 15\,148 silicon strip detector modules. Details on the pixel detector can be found in Ref.~\cite{Phase1Pixel}. For nonisolated particles of $1 < \pt < 10\GeV$ and $\abseta < 3.0$, the track resolutions are typically 1.5\% in \pt and 20--75\mum in the transverse impact parameter~\cite{DP-2020-049}. 

The CMS muon system consists of four types of gas-ionization detectors: drift tube chambers (DTs), cathode strip chambers (CSCs), resistive-plate chambers (RPCs), and triple-gas electron multiplier chambers (Triple-GEMs). The DT and CSC detectors are located in the regions of $\abseta < 1.2$ and $0.9 < \abseta < 2.4$, respectively, and are complemented by the RPCs in the range $\abseta < 1.9$. The Triple-GEMs are located in the region of $1.55 < \abseta < 2.18$. The chambers are arranged to maximize the coverage and to provide some overlap wherever possible. In both the barrel and endcap regions, the chambers are grouped into four ``muon stations'', separated by the steel absorber of the flux-return yoke. A detailed description of these detectors, including the gas composition, operating voltage, and performance, is reported in Ref.~\cite{CMS:2018rym}. More detailed descriptions of the CMS detector, together with a definition of the coordinate system used and the relevant kinematic variables, can be found in Refs.~\cite{CMS:2008xjf,CMS:2023gfb}. 

Events of interest are selected using a two-tiered trigger system. The first level, composed of custom hardware processors, uses information from the calorimeters and muon detectors to select events at a rate of around 100\unit{kHz} within a fixed latency of about 4\mus~\cite{CMS:2020cmk}. The second level, known as the high-level trigger (HLT), consists of a farm of processors running a version of the full event reconstruction software optimized for fast processing, and reduces the event rate to around 5\unit{kHz} before data storage~\cite{CMS:2016ngn,CMS:2021yvr}.

\section{Muon reconstruction and identification}
\label{sec:muons}
Muon tracks are first reconstructed independently in the inner tracker and in the muon system using Kalman filtering techniques~\cite{Fruhwirth:1987fm}. In the muon system, ``standalone muons'' are reconstructed by combining the information from the DT, CSC, RPC, and Triple-GEM detectors. To this end, the signals from the individual detector planes within the DT and CSC muon stations are first combined into track segments, which are thereafter used as the input to a Kalman filtering algorithm, at which stage the RPC and GEM hits are included in the pattern recognition. In the inner tracker, tracks are reconstructed iteratively, with each iteration seeded by a different combination of pixel or strip detector hits, to ensure all relevant track topologies are reconstructed~\cite{CMS:2014pgm}. Dedicated iterations seeded by standalone muons maximize the efficiency of muon track reconstruction. 

Tracker tracks and standalone muons with compatible momentum, direction, and position in the transverse plane are combined into ``global muons'', and a combined fit to all associated tracker and muon system signals is performed to determine the parameters of the global track. The trajectories of muons with very low \pt are bent in the magnetic field to such a degree that they do not reach all stations of the muon system. To recover such muons, for which a standalone muon might not be reconstructed, tracker tracks are extrapolated to the muon segments and a loose geometrical matching to DT and CSC segments is performed. If at least one compatible segment is found, the track is classified as a ``tracker muon''. The need for the muon to reach at least one muon station limits the acceptance for low \pt muons to 3.5\GeV in the barrel region of the detector and down to 1\GeV in the endcap. The muon reconstruction and the resulting efficiencies and momentum resolution are described in more detail in Ref.~\cite{CMS:2018rym}.

The properties of the resulting muon tracks are used to further identify muons for use in physics analyses, with ID techniques and working points chosen to match the desired trade-off between efficiency and background rejection. For soft muons, the Run~2 soft-muon MVA is the most common choice when analyzing Run~2 data. This MVA uses a BDT that is trained using \textsc{AdaBoost}~\cite{schapire2013explaining} on a simulated sample of muons reconstructed as global muons and whose tracker tracks survive the ``high purity'' track selection~\cite{CMS:2014pgm}. The muons are also required to have $\pt > 4\GeV$ and $\abseta < 1.4$ and signal muons are required to originate from a \PBzs or \PBz decay. Despite specifically targeting these decays, the BDT also performs well for soft muons from other sources. 

Several cut-based IDs are also available to identify soft muons in CMS, specifically the ``soft muon ID'', ``loose muon ID'', and ``medium muon ID'' working points~\cite{CMS:2018rym}. CMS recently introduced an MVA for muon ID (``Muon MVA'') targeting muons with $\pt > 10\GeV$~\cite{CMS:2023dsu}, which serves as another point of comparison for the Run~3 soft-muon MVA described in this paper. 

\section{Training and validation samples}
\label{sec:samples}
For training and validation of the Run~3 soft-muon MVA, a large sample of muons originating from both light- and heavy-flavor decays is needed. A large ($\mathcal{O}(10^{9})$ events) Monte Carlo (MC) simulation sample of inclusive dilepton production has been produced by simulating nondiffractive quantum chromodynamics (QCD) processes using \PYTHIA~8.306~\cite{Sjostrand:2014zea} at leading order (LO). The masses and decay widths for hadrons, as well as mixing angles between neutral \PB mesons, have been adjusted to match experimental results, where necessary. The events are filtered for the presence of two muons or two electrons at the generator level to reduce the computing costs of the following simulation and reconstruction steps. To include muons from decays in flight of \PGp and \PK mesons in the filtering, these decays, usually included in the detector simulation, have been enabled in \PYTHIA. For other samples described below, multiple versions matching the changing detector conditions throughout the 2022 and 2023 data taking exist. The significant computing cost involved in the production of this large sample prohibits the production of multiple versions and only a single sample matching the detector conditions in early 2022 exists. It has been verified that the Run~3 soft-muon MVA trained on this sample nevertheless generalizes well to the remaining data collected in 2022 and 2023.

To evaluate the efficiency of different classifiers, \jpsimm events are used. The \jpsi production is simulated at LO using \PYTHIA~8.306 with a minimum \pt of the \jpsi of 8\GeV. The background rate is measured in events containing high-\pt electrons, which are used to select the events independently of the pion and kaon decays constituting the background. They originate mostly from Drell--Yan (DY), \Wjets, and top quark pair production (\ttbar). The DY and \Wjets events are generated at next-to-LO (NLO) in QCD using \MGvATNLO~\cite{Alwall:2014hca}~v2.9.9. To maximize the number of events available for this study, a LO DY sample produced with \MGvATNLO~v2.6.5 is also used. The MLM~\cite{Alwall:2007fs} (FxFx~\cite{Frederix:2012ps}) scheme is used to match jets from the matrix element calculations and parton shower for the LO (NLO) samples. The \POWHEG 2.0~\cite{Alioli:2010xd,Frixione:2007vw,Nason:2004rx,Monni:2019whf} generator is used to simulate \ttbar production and additional $\mathrm{DY} \to \Pe\Pe$ events at NLO in QCD.

For all samples, the NNPDF 3.1 next-to-NLO (NNLO) parton distribution function set~\cite{Ball:2017pdf} is used, and parton showering and hadronization are simulated with \PYTHIA using the underlying event tune CP5~\cite{Sirunyan:2019dfx}. The response of the CMS detector is simulated using \GEANTfour 10.07~\cite{geant4} and the events are reconstructed with the same algorithms as those used for the data. The effect of pileup is modeled by superimposing the primary interaction with additional minimum bias interactions simulated with \PYTHIA.

Collision data collected in 2022 and 2023, corresponding to an integrated luminosity of 62\fbinv, are used to validate the performance of the classifier. To measure the selection efficiency, events were collected with triggers that require at least one muon (``single-muon triggers'') without imposing isolation requirements, with minimum \pt thresholds ranging from 8 to 21\GeV. These triggers were prescaled, \ie, only a certain fraction of accepted events were recorded, with the prescale factor decreasing with increasing \pt. To increase the size of the data sample, additional events from the same years were collected using an unprescaled trigger requiring at least one isolated muon with $\pt > 24\GeV$~\cite{CMS:2021yvr}. The background rate is measured in a sample of events collected using a trigger requiring the presence of at least one electron (``single-electron trigger'') with $\pt > 30\GeV$.

\section{Model training and validation}
\label{sec:models}
The Run~3 soft-muon MVA is a BDT classifier trained using gradient boosting~\cite{friedman2001greedy}, implemented with \textsc{XGBoost}~\cite{Chen:2016:XST:2939672.2939785} via the \textsc{scikit-learn} package~\cite{scikit-learn}. The parameters of the model were optimized using the grid search approach to maximize the area under the receiver operating characteristic (ROC) curve (AUROCC) achieved during training.

Alternative models, such as histogram-based gradient boosting (HGB) or deep neural networks (DNN), were similarly optimized and trained. The ROC curves for all models were compared. The \textsc{XGBoost} and HGB models performed similarly, and better than the DNN. In the end, the \textsc{XGBoost} model was chosen for its ease of use and because it could achieve good performance with fewer input features compared to the HGB model.  

Since the Run~3 soft-muon MVA aims to improve acceptance for soft muons with very small \pt and high \abseta, the preselection applied to the muons used for training is kept minimal. A loose kinematic selection of $\pt > 2\GeV$ and $\abs{\eta} < 2.4$ is applied and the muons are required to be reconstructed as tracker or global muons. For the training and validation, muons are labeled as signal when they originate from the decay of heavy-flavor hadrons, and as background when they originate from the decay in flight of pions and kaons. Muons from other sources are not considered. This definition results in significant differences in the \pt and $\eta$ distributions of signal and background muons, with signal muons exhibiting a harder \pt spectrum. Weights are therefore derived from the ratio of the \pt distributions of signal and background muons in bins of $\eta$, and the sample of background muons is corrected to reproduce the \pt distribution of the signal muons.  

The decays of \PGp and \PK mesons into muons introduces a discontinuity in the trajectory observed in the tracker. Consequently, background muons are expected to have fewer hits associated with their inner track compared to signal muons and a poor match of the inner track to hits in the muon system. A total of 26 input features to the BDT are used, which are sensitive to these differences between signal and background muons. Among them are the $\eta$ and \pt of the muon and the number of hits in the pixel detector, strip detector layers with measurements, and muon system hits associated with the muon. The number of ``lost hits'', \ie, tracker layers where a hit is expected to be associated with the inner track but none is found, is counted separately for detector layers before and after the first and last hit on the track (as seen from the center of the detector outward), and for layers in between them. Similarly, the fraction of valid hits associated with the tracker track of the muon compared with the expected maximum is considered. In case the muon is a global muon, the $\chi^2$, normalized by the number of degrees of freedom, as well as the negative logarithm of the $p$-value of the global track fit and the $\chi^2$ of the spatial matching between the inner and standalone tracks are considered. Additional inputs are the ``high purity'' flag of the inner track and the maximum $\chi^2$ probability that the inner track consists of two independent tracks. Finally, if the muon has associated track segments in the first or second muon stations, a set of up to 12 variables related to the quality of the matching of the inner track to these segments is used. For these, the inner track is projected to the muon station, and its position and direction are compared to the muon segment along different axes.  Three positional and three directional variables are defined in each station, taking into account the measurement uncertainties. If a feature is not available for a muon, it is set to an unphysical default value.

\begin{figure}[htb]
    \centering
    \includegraphics[width=0.49\textwidth]{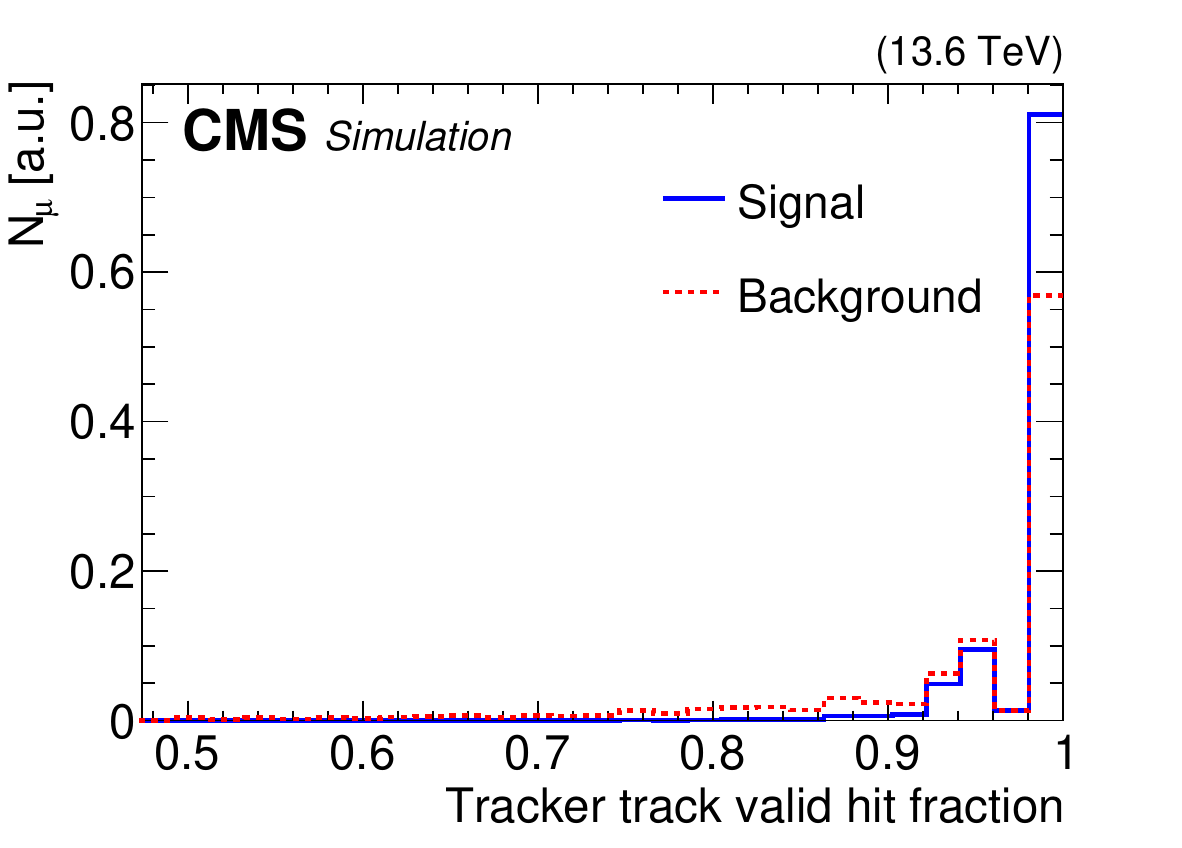}
    \includegraphics[width=0.49\textwidth]{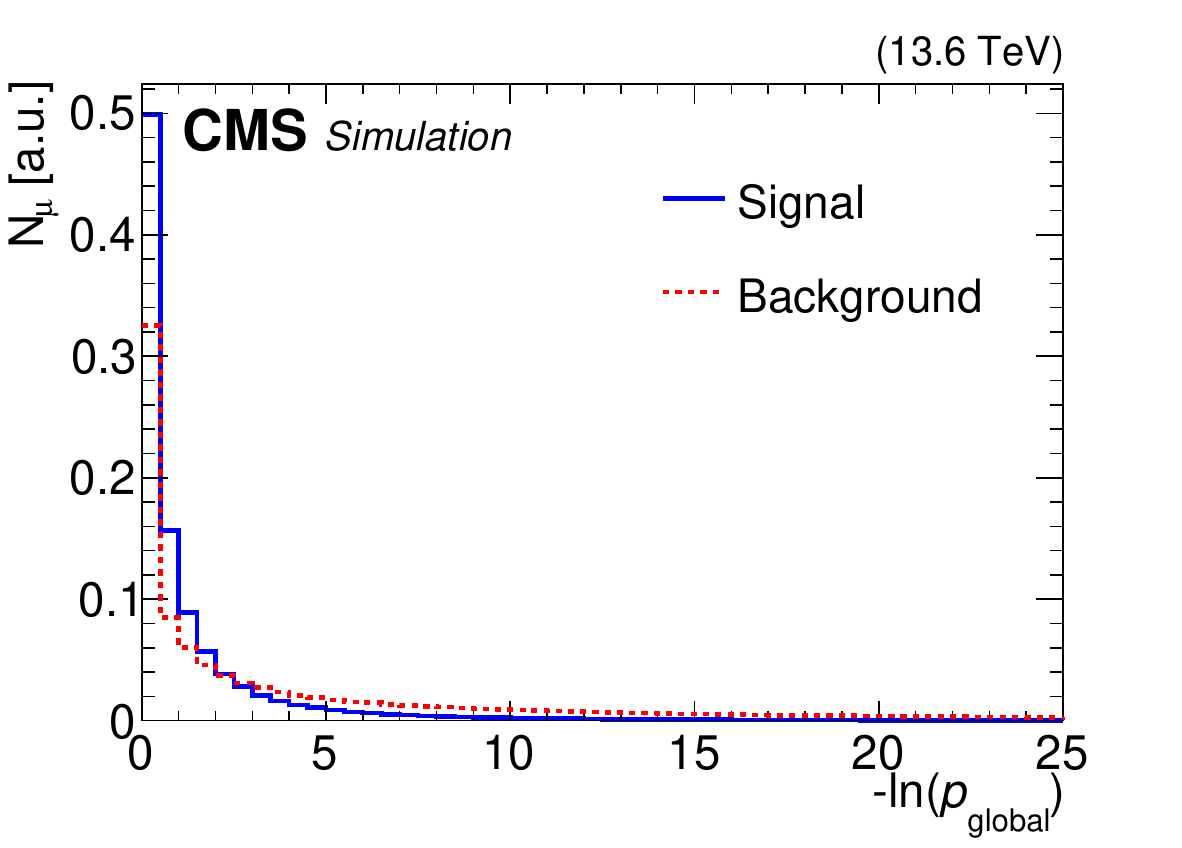}
    \caption{The distribution of the fraction of valid hits on the tracker track (left) and the $-\ln(p_{\text{global}})$ (right), where $p_{\text{global}}$ is the global track fit probability, for signal muons (solid blue line) and background muons (dashed red line) used in the training.}
    \label{fig:var_compare}
\end{figure}

Among all the input variables, the fraction of valid hits on the tracker track and the global track fit probability have the strongest distinguishing power. Their distributions for signal and background muons, normalized to the same area, are shown in Fig.~\ref{fig:var_compare}.

In comparison, the Run~2 soft-muon MVA was trained using 14 input features, largely overlapping with the ones chosen for the Run~3 version. Of the newly included features in the Run~3 soft-muon MVA, the more detailed information about the hit pattern in the inner tracker and the information about the compatibility of the inner track with the muon segments contribute the most to the improvement in the performance.

\begin{figure}[ht]
    \centering 
    \includegraphics[width=0.4\textwidth]{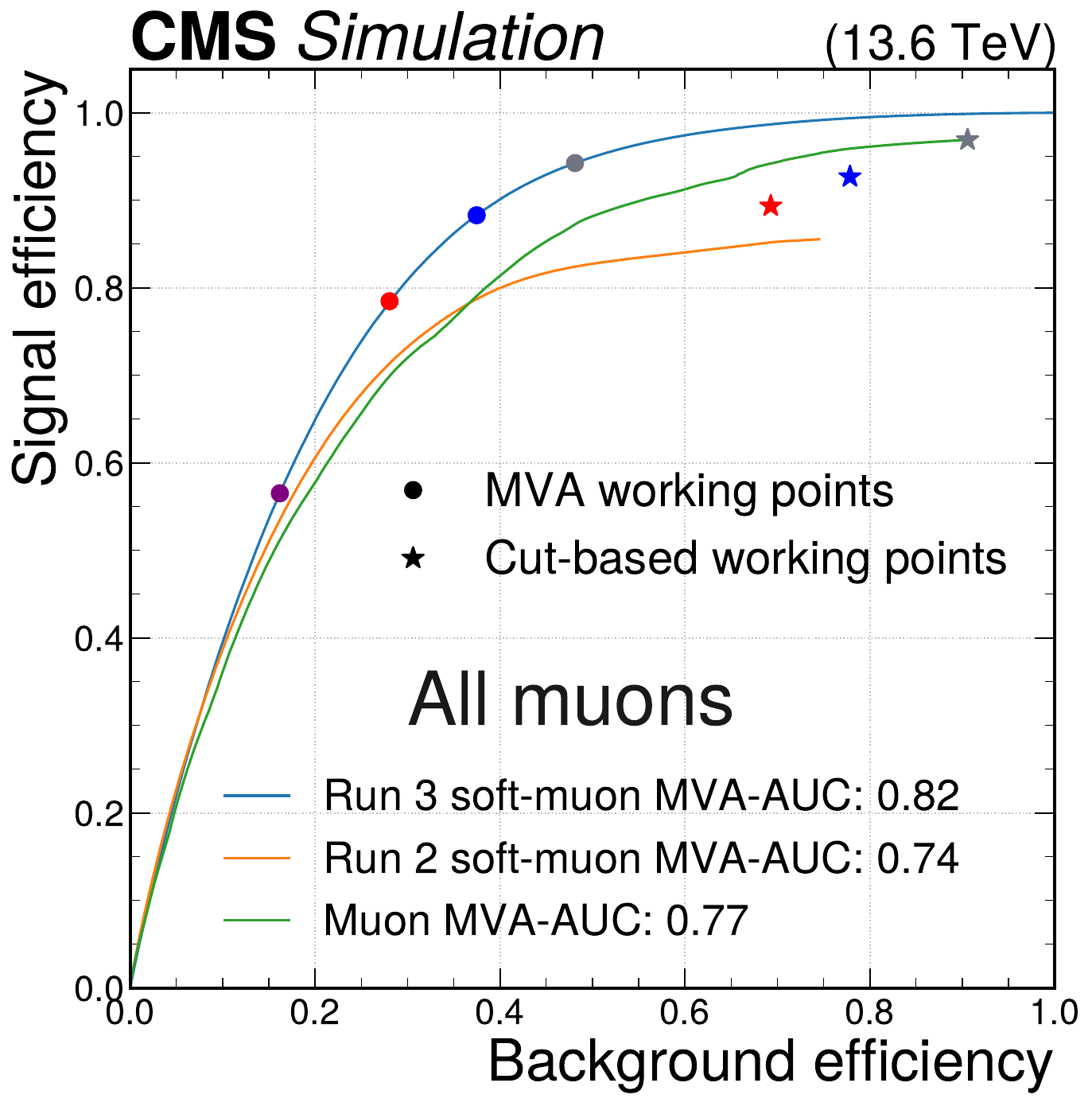}\\
    \includegraphics[width=0.4\textwidth]{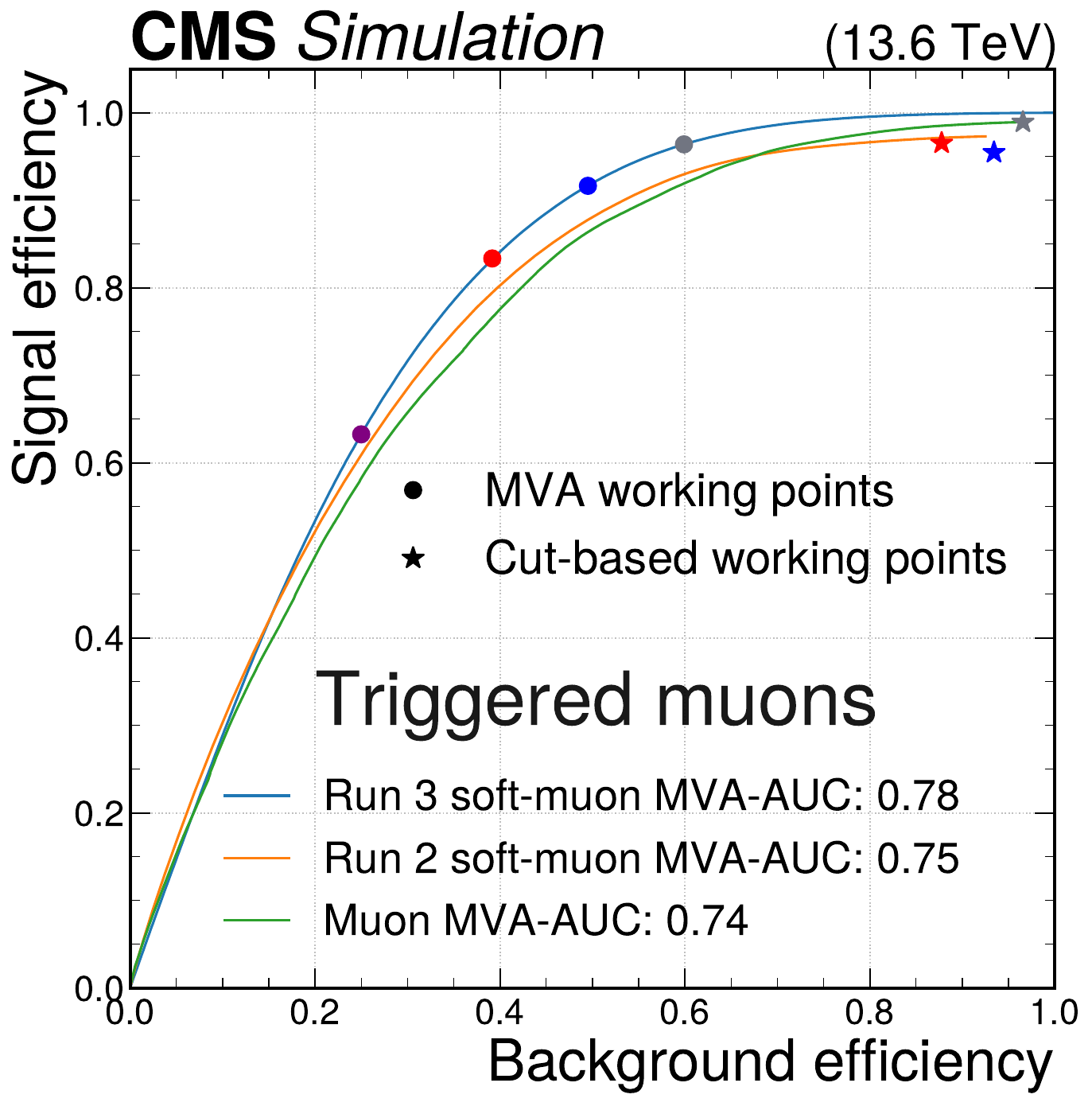}
    \includegraphics[width=0.4\textwidth]{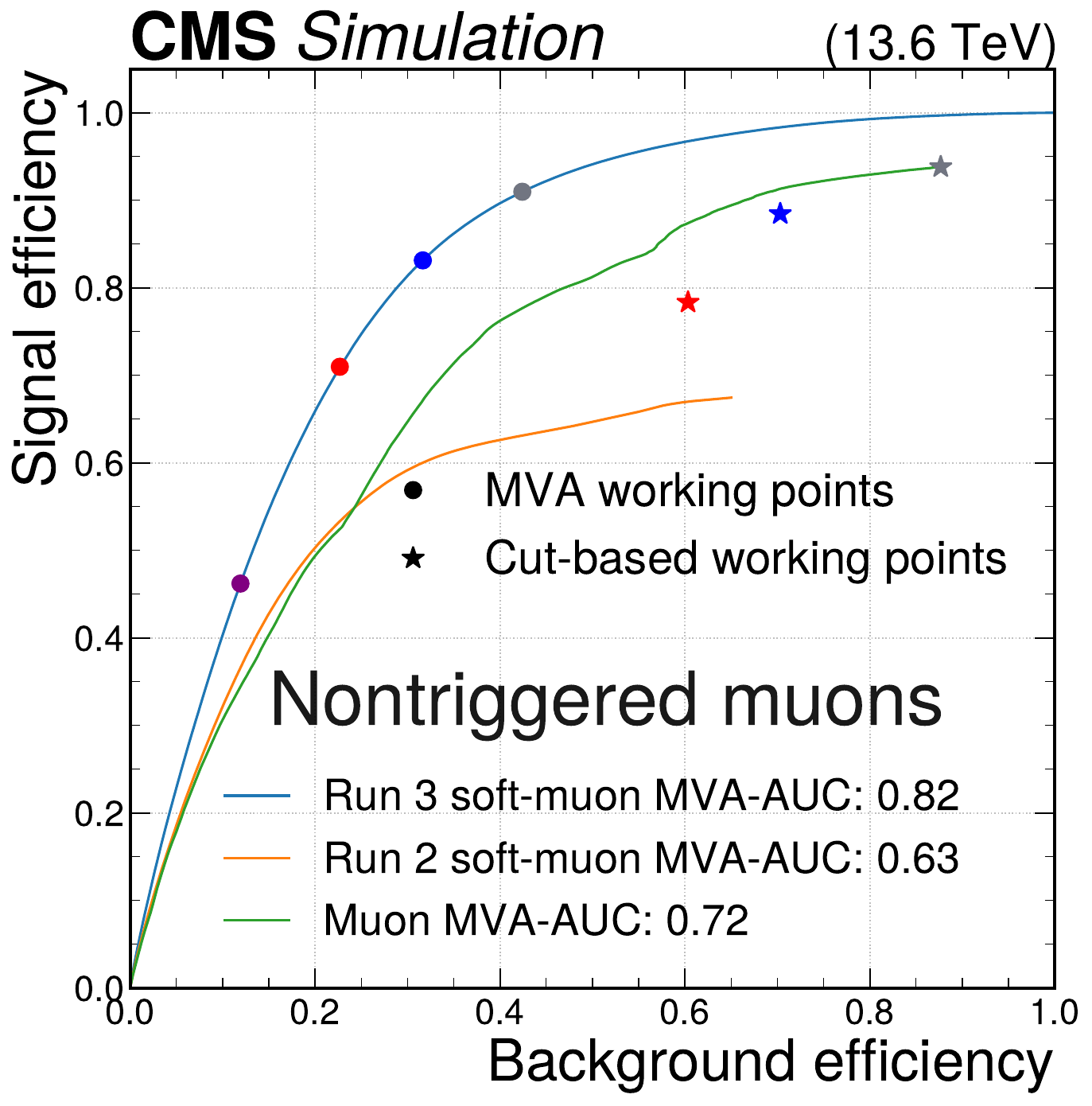}    
    \caption{The ROC curve for the Run~3 soft-muon MVA (blue line) compared to those for the Run~2 soft-muon MVA (orange line) and the Muon MVA (green line). The working points defined in Table~\ref{tab:XGBWPs} are indicated with round, colored markers, with the very loose, loose, medium, and tight working points being represented by the grey, blue, red, and purple markers, respectively. For comparison, the performance of the cut-based IDs is indicated by the colored stars, with the loose, soft, and medium IDs being represented by the grey, blue, and red markers, respectively. The upper plot shows the ROC curves for all muons, while the lower ones split the muon sample into those that fired the HLT (left) and those that did not (right).}
    \label{fig:auroc}
\end{figure}

The training is performed using about 10 million muons from the simulated inclusive dilepton sample, and another 5 million muons are used for validation. Of these muons, roughly two thirds are signal and the remainder are background muons. A maximum of 5000 boost rounds is set and the training is configured to maximize the AUROCC. The resulting ROC curve for the Run~3 soft-muon MVA is shown in the upper plot of Fig.~\ref{fig:auroc}, where it is compared with the Run~2 soft-muon MVA and the Muon MVA, as well as several cut-based IDs. For the Run~3 version of the soft-muon MVA, the working points defined below are indicated. The Run~2 soft-muon MVA and the Muon MVA use tighter preselections in their training compared with the Run~3 soft-muon MVA, which prevents them from being evaluated in the full phase space considered here, resulting in their ROC curves not reaching full signal and background efficiency. To calculate the AUROCC in these cases, the curves are linearly extrapolated to (1,1). The Run~3 soft-muon MVA achieves an AUROCC value of 0.82 and clearly outperforms both the Run~2 version (0.74) and the Muon MVA (0.77) over the full range of signal and background efficiency combinations. Applying the stricter preselection of the Run~2 soft-muon MVA, the AUCROCC values change to 0.81, 0.79, and 0.77, respectively, illustrating that the Run~3 soft-muon MVA improves upon the previous iteration also in this phase space. The Muon MVA performs comparably to the Run~2 soft-muon MVA, despite not being optimized for $\pt < 10\GeV$ since it uses a looser preselection that more closely matches the preselection used in the training of the Run~3 soft-muon MVA. All MVAs significantly outperform the cut-based IDs, offering significantly better background rejection at the same signal efficiency. 

For events to pass the HLT, muons are usually required to pass certain quality and \pt criteria already at trigger level~\cite{CMS:2021yvr}. Requiring muons to have fired the trigger (``triggered muons'') therefore results in a sample enriched in signal muons with higher \pt compared to those that fail it (``nontriggered muons''). Both classes are roughly evenly represented in the training and validation samples. To compare the performance of the different MVAs, the two lower plots in Fig.~\ref{fig:auroc} show the ROC curves separately for triggered and nontriggered muons. The Run~3 soft-muon MVA exhibits similar performance in both cases, while the other two MVAs perform significantly better on triggered muons. We find that the new Run~3 soft-muon MVA improves most upon the previous classifiers for triggered muons with $\pt < 4\GeV$ and for nontriggered muons for all \pt. For triggered muons with $\pt > 4\GeV$, only a modest improvement for the Run~3 soft-muon MVA over the Run~2 version is observed.

\begin{figure}
    \centering
    \includegraphics[width=0.5\linewidth]{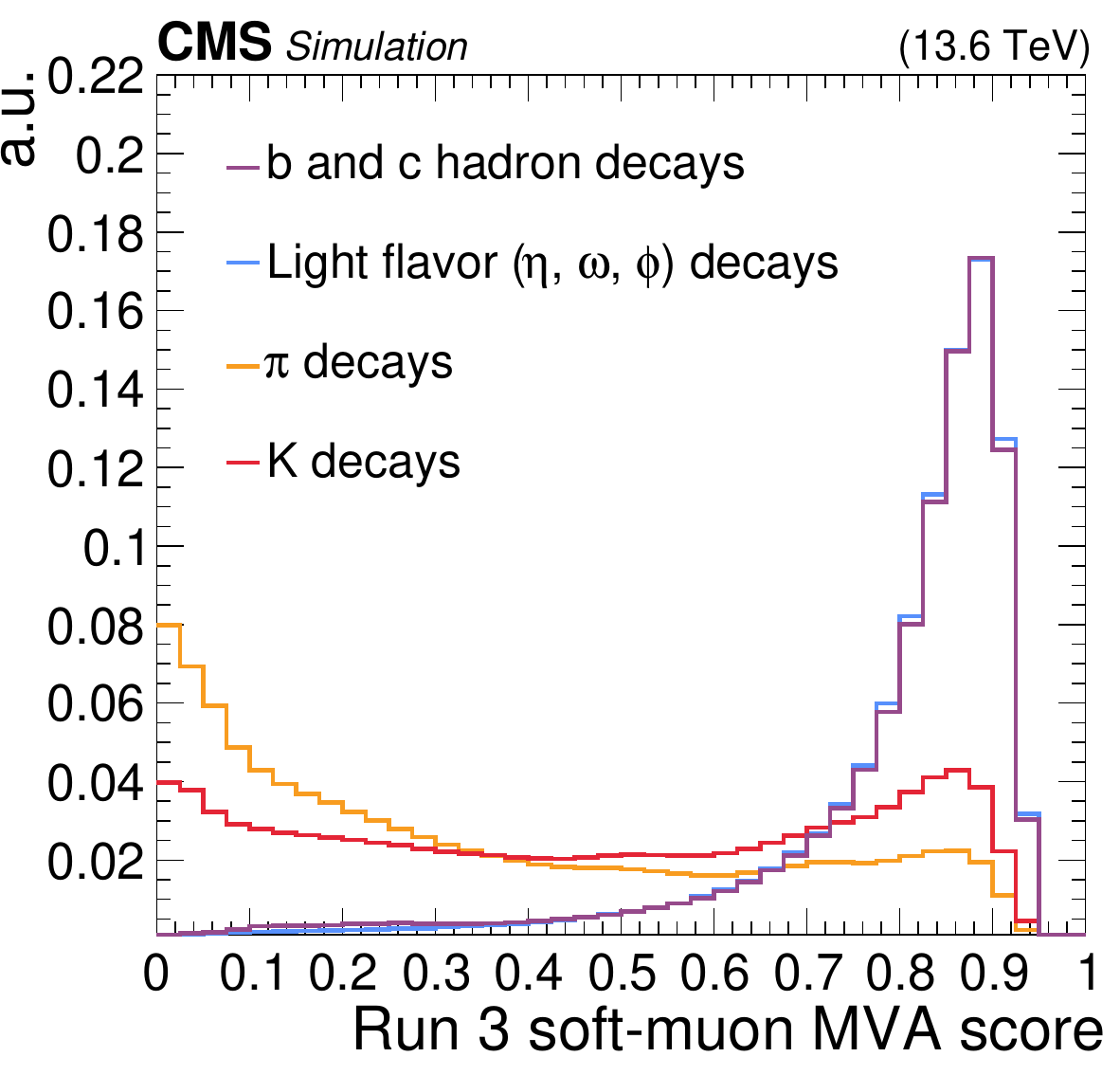}
    \caption{Distribution of the Run~3 soft-muon MVA score for muons of different origins.}
    \label{fig:score}
\end{figure}

To illustrate the expected performance of the Run~3 soft-muon MVA classifier for muons from different sources, the distribution of the MVA score is shown in Fig.~\ref{fig:score}. Muons from the inclusive dilepton sample are used, and reconstructed muons are matched to simulated ones based on detector hits to categorize them as originating from the decay of hadrons containing \PQb and \PQc quarks; \PGh, \PGo, and \PGf mesons; and \PGp and \PK mesons. The score distribution is strongly peaked towards maximal values for signal muons from heavy-flavor decays. This is also true for muons from the decays of short-lived light-flavor mesons, such as \PGh, \PGo, and \PGf mesons, despite not being included in the signal or background definitions used in the training of the MVA. The score for background muons from \PGp meson decays peaks at zero but has a long tail all the way to the highest score values. As \PK mesons have a shorter lifetime compared to \PGp mesons, muons from their decays are less easily distinguished from signal muons, and their MVA score is roughly evenly distributed, with slight peaks both towards zero and the maximal score value. Overall, it is evident that the Run~3 soft-muon MVA is well suited to select muons from the decay of objects with a wide range of invariant masses and to efficiently suppress \PGp and \PK meson decays. The MVA has also been verified to have similar performance for muons from \tautrimu decays, which are not present in this sample of simulated events.

Four benchmark working points have been defined, namely, ``very loose'', ``loose'', ``medium'', and ``tight'', according to the desired trade-off between the efficiency and background rejection. The medium working point is chosen to have a comparable background rejection as the working point for the Run~2 soft-muon MVA used in Ref.~\cite{CMS:2019bbr}. The other working points are chosen so that the background rejection increases by roughly a factor of 2 when comparing it to the next tighter one. The resulting thresholds on the classifier score are summarized in Table~\ref{tab:XGBWPs}, together with performance metrics evaluated on the inclusive dilepton sample used in the training of the classifier. The performance in collision data is evaluated separately in Section~\ref{sec:performance}. The efficiency is defined as the fraction of muons from the targeted heavy-flavor decays that are selected. It is above 95\% for the very loose working point and decreases to 58\% for the tight working point. The purity of the selected muon sample, defined as the fraction of the selected muons originating from heavy-flavor decays, increases from 69 (84)\% to 81 (90)\% between the very loose and tight working points for muons with $\pt > 2\,(4)\GeV$. The background to the selected sample consists overwhelmingly of muons from light-flavor decays. Contributions from other sources, such as hadrons whose tracks are mistakenly matched to muon detector signals or detector noise misidentified as muons comprise 4--7\% of the selected muon candidates, depending on the working point, for muons with $\pt >2\GeV$. This contribution falls quickly with \pt and is 1--2\% for $\pt > 4\GeV$. Hadron punch-through, where hadronic showers reach the muon detector, is found to be negligible for all working points in the inclusive dilepton sample and is likewise negligible for working points tighter than medium in fully hadronic topologies.

\begin{table}[htb]
\centering
 \topcaption{Working points for the Run~3 soft-muon MVA classifier. For each working point, the efficiency of selecting muons from heavy-flavor decays, the purity of the selected muon sample, and the fraction of selected objects that are not associated with a real muon are given. These metrics are evaluated on the inclusive dilepton sample.}
 \cmsTable{
 \begin{tabular}{lcccc}
    Name        & MVA score & Efficiency    & Purity $\pt > 2\,(4)\GeV$ & Non-muon fraction $\pt > 2\,(4)\GeV$  \\ \hline
    Very loose  & ${>}0.47$ & 95\%          & 69 (84)\%                 & 7.0 (2.1)\%                           \\
    Loose       & ${>}0.63$ & 89\%          & 73 (87)\%                 & 5.9 (1.5)\%                           \\
    Medium      & ${>}0.74$ & 80\%          & 76 (88)\%                 & 4.7 (1.3)\%                           \\
    Tight       & ${>}0.83$ & 58\%          & 81 (90)\%                 & 3.6 (1.2)\%
 \end{tabular}
 }
 \label{tab:XGBWPs}
\end{table}

\section{Performance in data and simulation}
\label{sec:performance}
To verify that the Run~3 soft-muon MVA performance in collision data matches the expectation from simulation, it is evaluated in both cases using multiple metrics. The efficiency of the classifier, \ie, the fraction of signal muons passing the selection, is the first relevant metric to consider and is described in Section~\ref{sec:eff}. Then, the background rate, \ie, the fraction of background muons passing the selection, is described in Section~\ref{sec:bkgRate}.

\subsection{Efficiency}\label{sec:eff}
The efficiency of selecting signal muons is evaluated using a ``tag-and-probe'' method similar to the one described in Ref.~\cite{CMS:2010svw}, but using \jpsimm decays. One of the muons, the ``tag'', is required to have $\pt > 8\GeV$ and $\abseta < 2.4$, pass the tight muon ID criteria~\cite{CMS:2018rym}, and to have passed one of the single-muon triggers described above. The other muon, the ``probe'', is required to have $\pt > 2\GeV$ and $\abseta < 2.4$ and to originate from within 0.5\cm of the primary vertex (\ie, the vertex corresponding to the hardest scattering in the event, evaluated using tracking information alone, as described in Section 9.4.1 of Ref.~\cite{CMS-TDR-15-02}.) in the direction along the beam line. It is also required to be reconstructed as a tracker or global muon, matching the preselection for the training of the classifier. The two muons are further required to have opposite electric charge, an angular separation $\DR \equiv \sqrt{\smash[b]{(\Delta\eta)\,^2 + (\Delta\phi)\,^2}} > 0.3$ and an invariant mass in the range $2.7 < m_{\Pgm\Pgm} < 3.5\GeV$ around the nominal \jpsi meson mass.

The efficiency is defined as the fraction of probe muons that pass the muon ID criterion in question. The selection outlined above accepts not only \jpsi events but also a significant continuous background that needs to be subtracted to ensure an unbiased measurement. The invariant mass distributions for events with probe muons that pass or fail the muon ID are fitted separately with a probability density function that consists of models for the \jpsi peak and the background. The \jpsi model is a convolution of a histogram template of the \jpsi peak at the generator level, taken from the \PYTHIA simulation, and a Gaussian function to model the detector resolution. The background shape is a combination of an error function with an exponential decay. Systematic uncertainties related to the choice of fit functions and the range and binning of the mass distribution are evaluated and found to be small.

\begin{figure}
    \centering
    \includegraphics[width=0.5\linewidth]{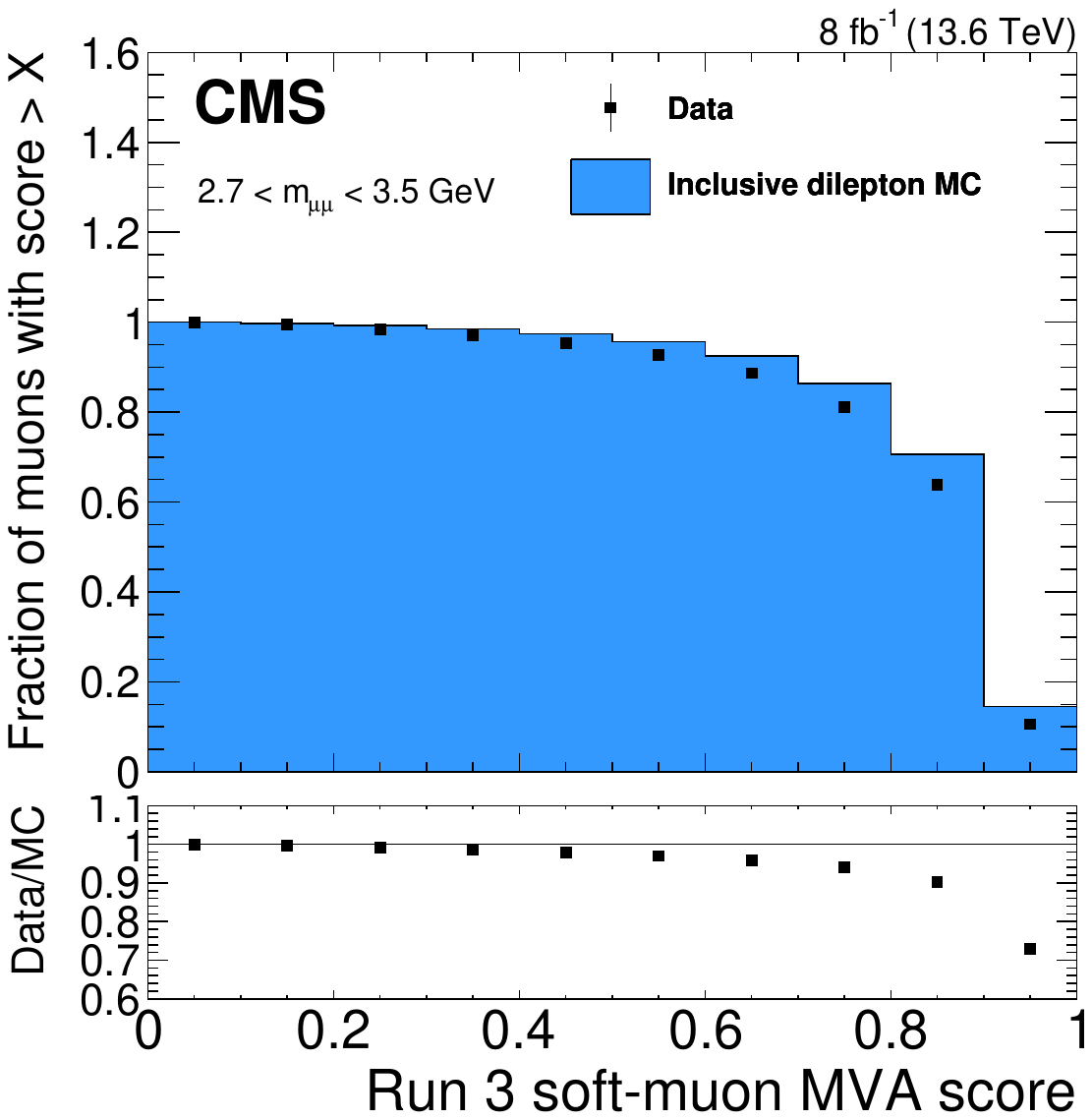}
    \caption{Cumulative distribution of the classifier score of the Run~3 soft-muon MVA, comparing data from early 2022 (black markers) with inclusive dilepton simulation (blue histogram) for muon pairs with $ 2.7 < m_{\Pgm\Pgm} < 3.5\GeV$. The lower panel shows the ratio of data to simulation. Uncertainties represented by the vertical bars are statistical only.}
    \label{fig:dataMC}
\end{figure}

A comparison of the distribution of the score of the Run~3 soft-muon MVA, integrated above a threshold, between data and MC simulation after applying this selection is shown in Fig.~\ref{fig:dataMC}. The inclusive dilepton MC sample used in the training of the MVA is employed since it better describes the background contribution to this selection. As described above, this sample is available only in a version matching the data-taking conditions in early 2022. Thus, only this subset of data is shown. Agreement between data and simulation is observed over a large range of score values, with significant deviations for very large values of the score, which are relatively abundant and correctable in simulation. These deviations predominantly affect muons detected in the detector endcaps and result from the inability to precisely model the detailed detector response. Input features sensitive to the precise alignment of the detector are particularly affected.

\begin{figure}
    \centering
    \includegraphics[width=0.49\textwidth]{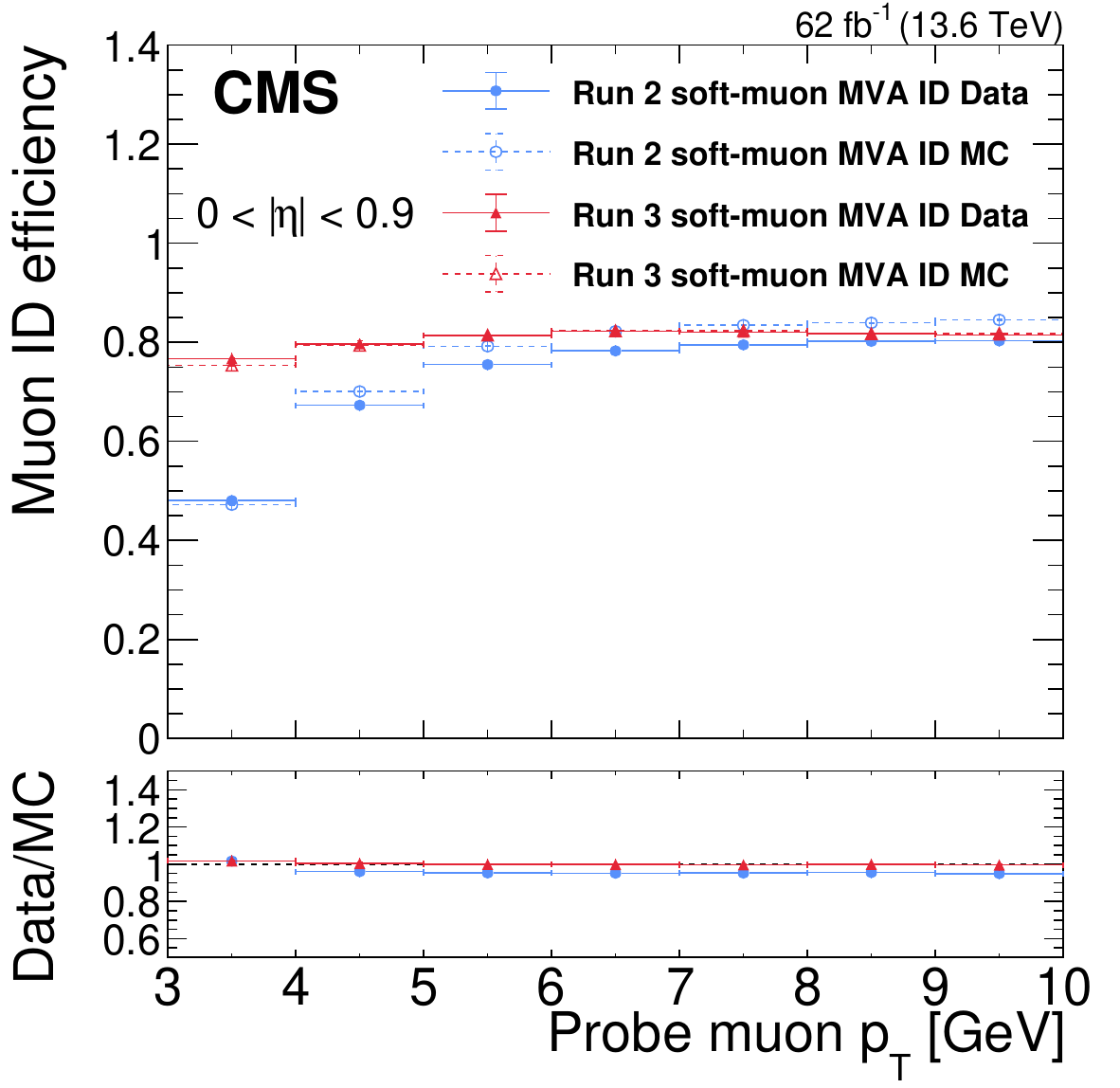}
    \includegraphics[width=0.49\textwidth]{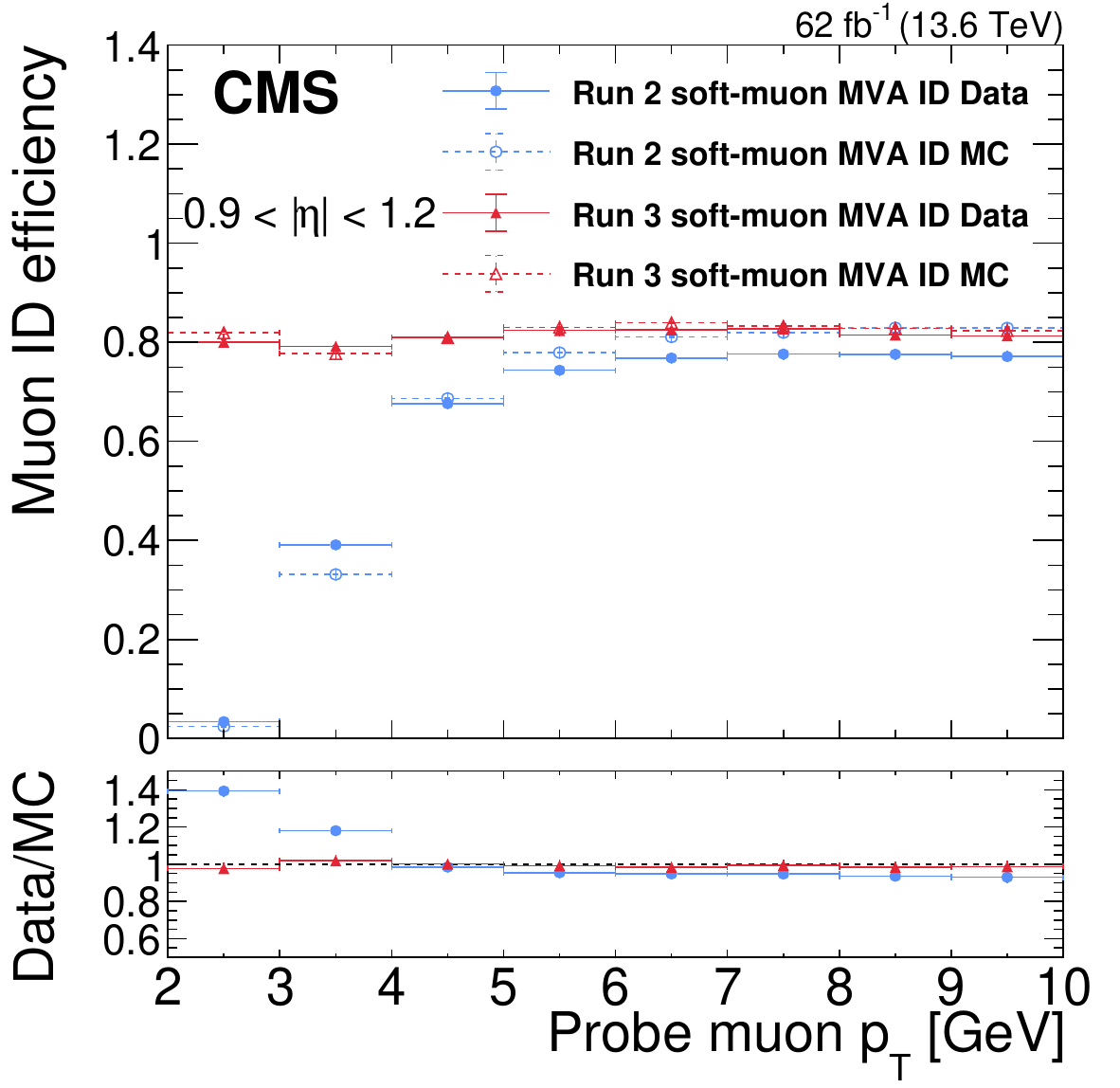}
    \includegraphics[width=0.49\textwidth]{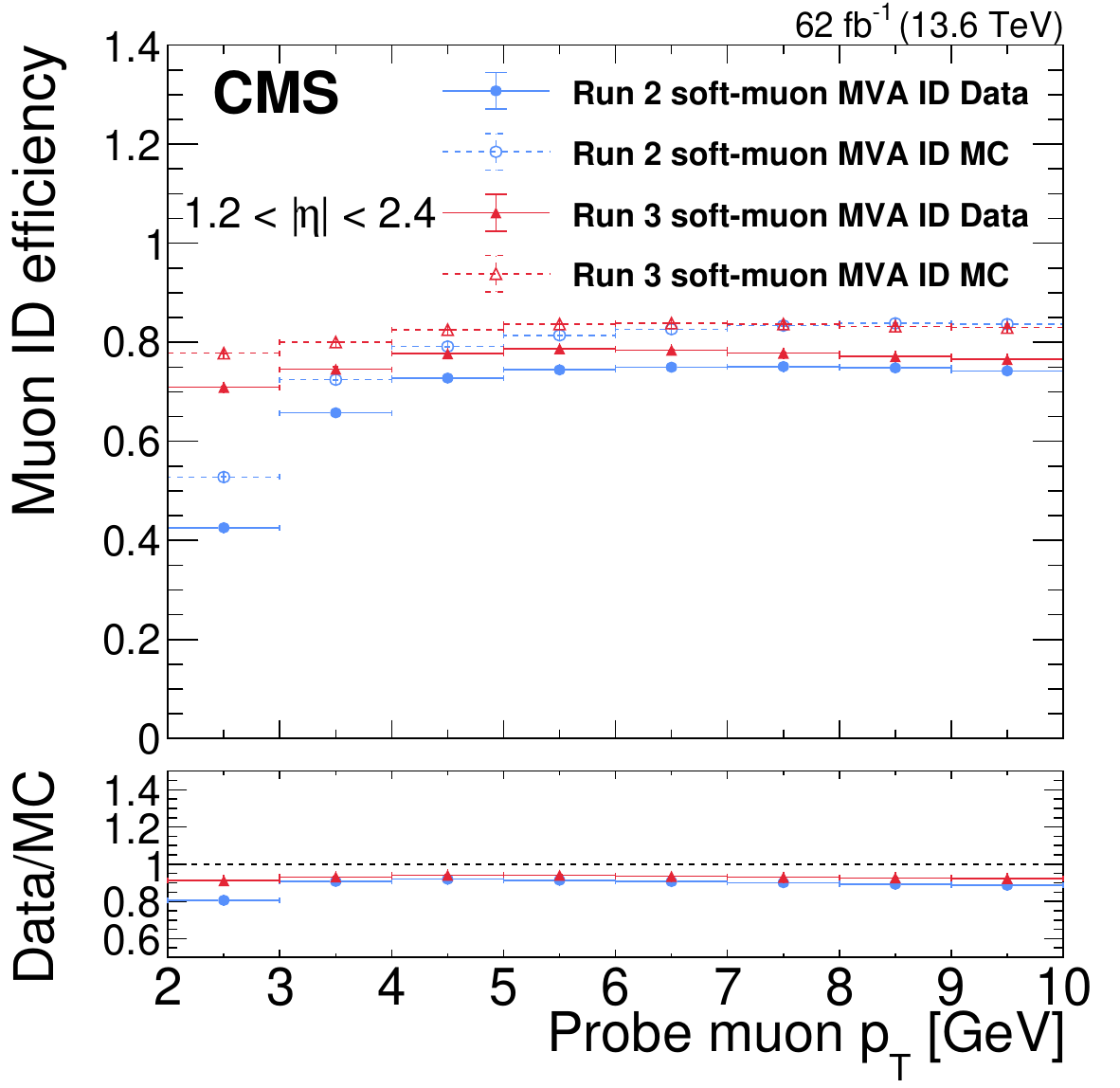}    
    \caption{Efficiencies of the Run~2 (blue) and Run~3 (red) soft-muon MVA as functions of muon \pt for muon $0 < \abseta < 0.9$ (upper left), $0.9 < \abseta < 1.2$ (upper right), $1.2 < \abseta < 2.4$ (lower). For the Run~3 soft-muon MVA, the medium working point is used. The vertical bars indicate the total uncertainty including statistical and systematic uncertainties.}
    \label{fig:effpt}
\end{figure}

\begin{figure}
    \centering
    \includegraphics[width=0.49\textwidth]{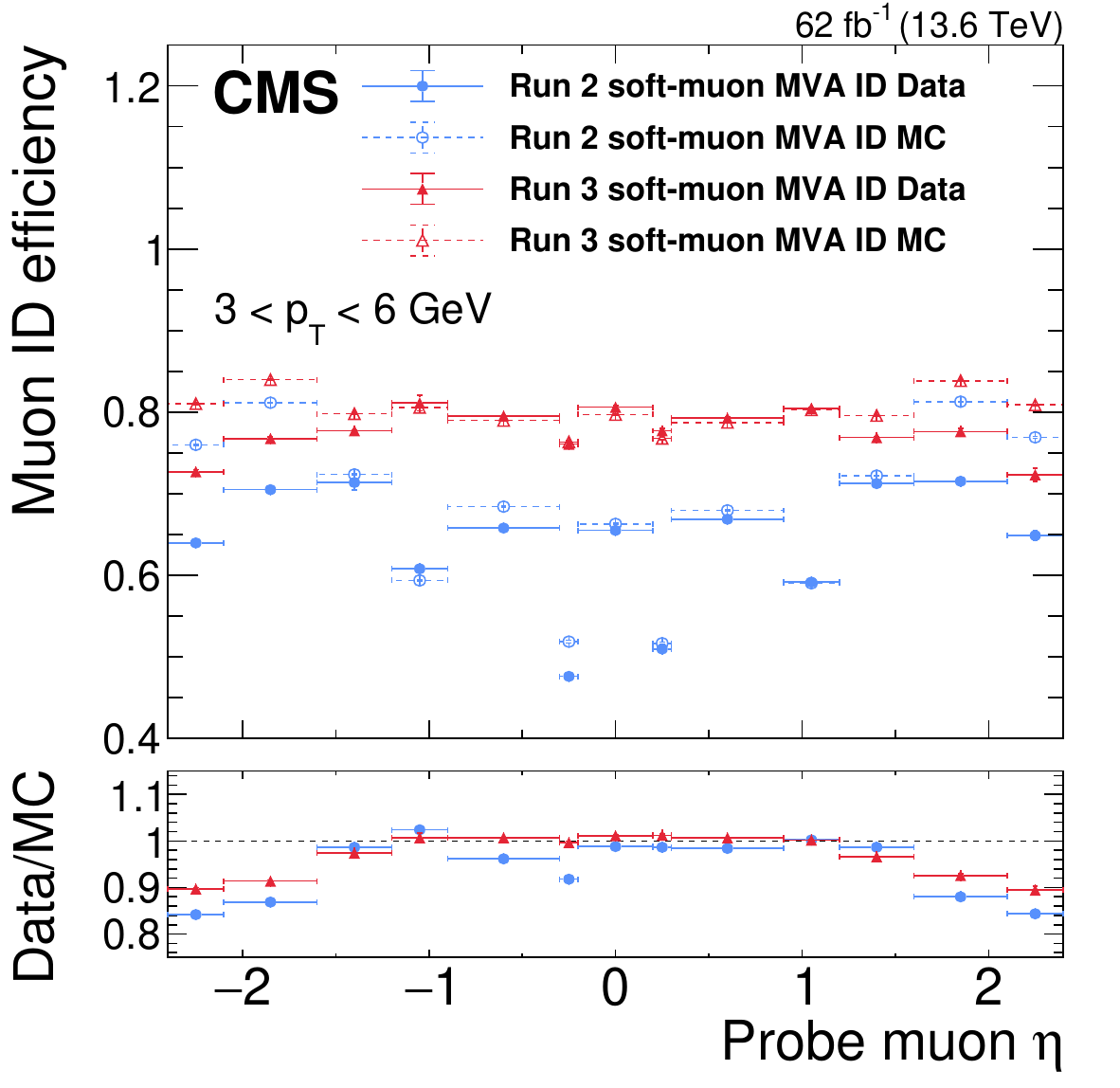}  
    \includegraphics[width=0.49\textwidth]{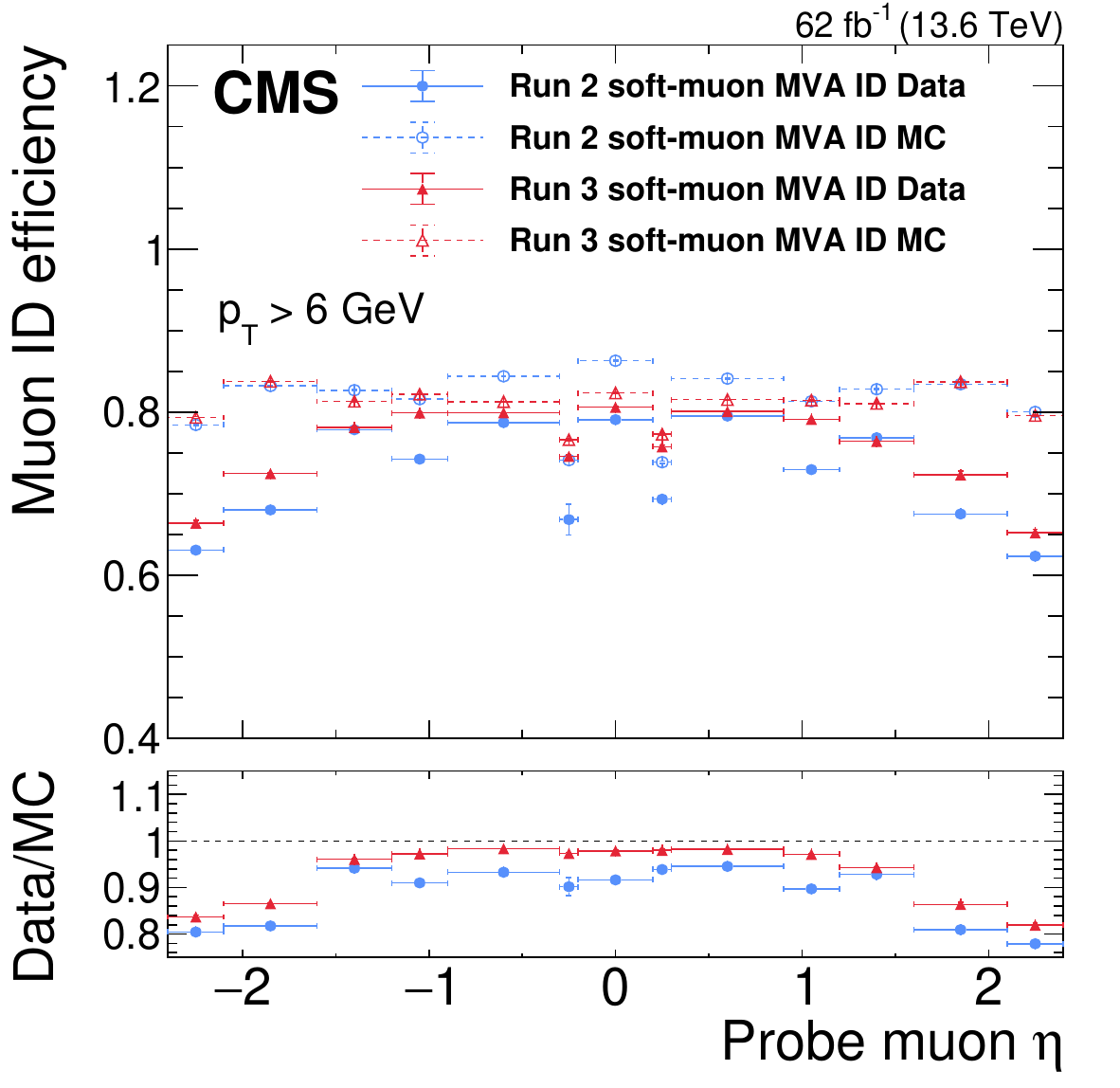}   
    \caption{Efficiencies of the Run~2 (blue) and Run~3 (red) soft-muon MVA as functions of muon $\eta$ for muons with $3 < \pt < 6\GeV$ (left) and $\pt > 6 \GeV$ (right). For the Run~3 soft-muon MVA, the medium working point is use. The vertical bars indicate the total uncertainty including statistical and systematic uncertainties.}
    \label{fig:effeta}
\end{figure}

The efficiency for the medium working point of the Run~3 soft-muon MVA is shown in Fig.~\ref{fig:effpt} as a function of muon \pt in several bins of muon \abseta, and in Fig.~\ref{fig:effeta} as a function of muon $\eta$ in two bins of muon \pt. In contrast to Fig.~\ref{fig:dataMC}, the efficiency measurements are performed using the full 2022 and 2023 data sets and the \jpsi samples are used to evaluate the efficiency in simulation. As muons with $\pt < 3\GeV$ do not reach the muon system in the central part of the detector, the efficiency as a function of \pt is shown for $\pt>3\GeV$ when $\abseta < 0.9$ and for $\pt > 2\GeV$ for larger values of \abseta. The efficiency as a function of muon $\eta$ reflects the physical structure of the CMS muon system; for example, the efficiency losses around $\abseta \sim 0.2-0.3$ are caused by gaps between the wheels~\cite{CMS:2018rym}. The efficiency of the Run~2 soft-muon MVA is shown for comparison, as its working point has a similar background rate as the medium working point for the Run~3 version. The Run~3 soft-muon MVA is significantly more efficient for muons with $\pt < 4\GeV$. This is the case especially for central muons, but the Run~3 soft-muon MVA is more efficient than its Run~2 counterpart in every $\eta$ and \pt bin. The performance of the Run~3 soft-muon MVA is better described by simulation than the Run~2 version. Significant differences of up to 18\% between the efficiency in data and simulation for the Run~3 MVA are observed only for $\abseta > 1.6$, caused by differences in data and simulation in the distribution of input features, especially for the variables describing the matching between the inner track and segments in the first two muon stations. Additionally, the data rates in this region in Run~3 are much higher than the CSC readout was designed to handle, leading to a fraction of CSC hits being lost before mitigation measures were introduced in 2023. This difference will be accounted for at the analysis level by adjusting the efficiency in simulation through the application of scale factors. The Run~3 soft-muon MVA shows a decrease of efficiency for higher \pt values, especially for muons with $\abseta > 1.2$, but still retains higher efficiency in data than the Run~2 MVA for \pt values up to 10\GeV. As the background contribution is much lower at higher \pt, the reduced efficiency in this phase space can easily be addressed at the analysis level by adjusting the classifier score threshold. The measured efficiencies were stable against the evolving detector conditions during data taking in 2022 and 2023, with the exception of the efficiency for forward muons, which improved after the loss of CSC hits in high pileup conditions was mitigated during 2023.

\begin{figure}
    \centering
    \includegraphics[width=0.49\textwidth]{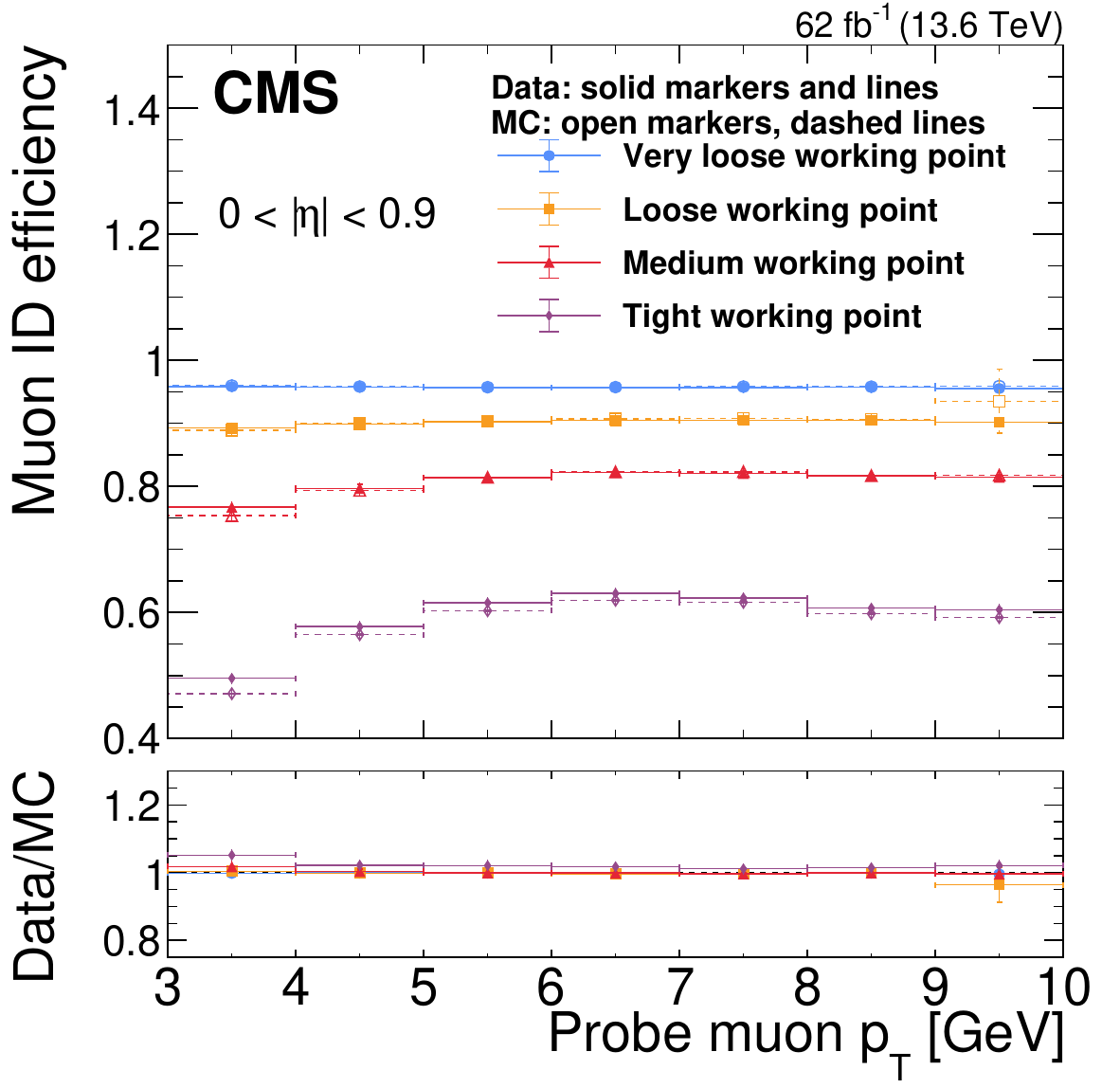}
    \includegraphics[width=0.49\textwidth]{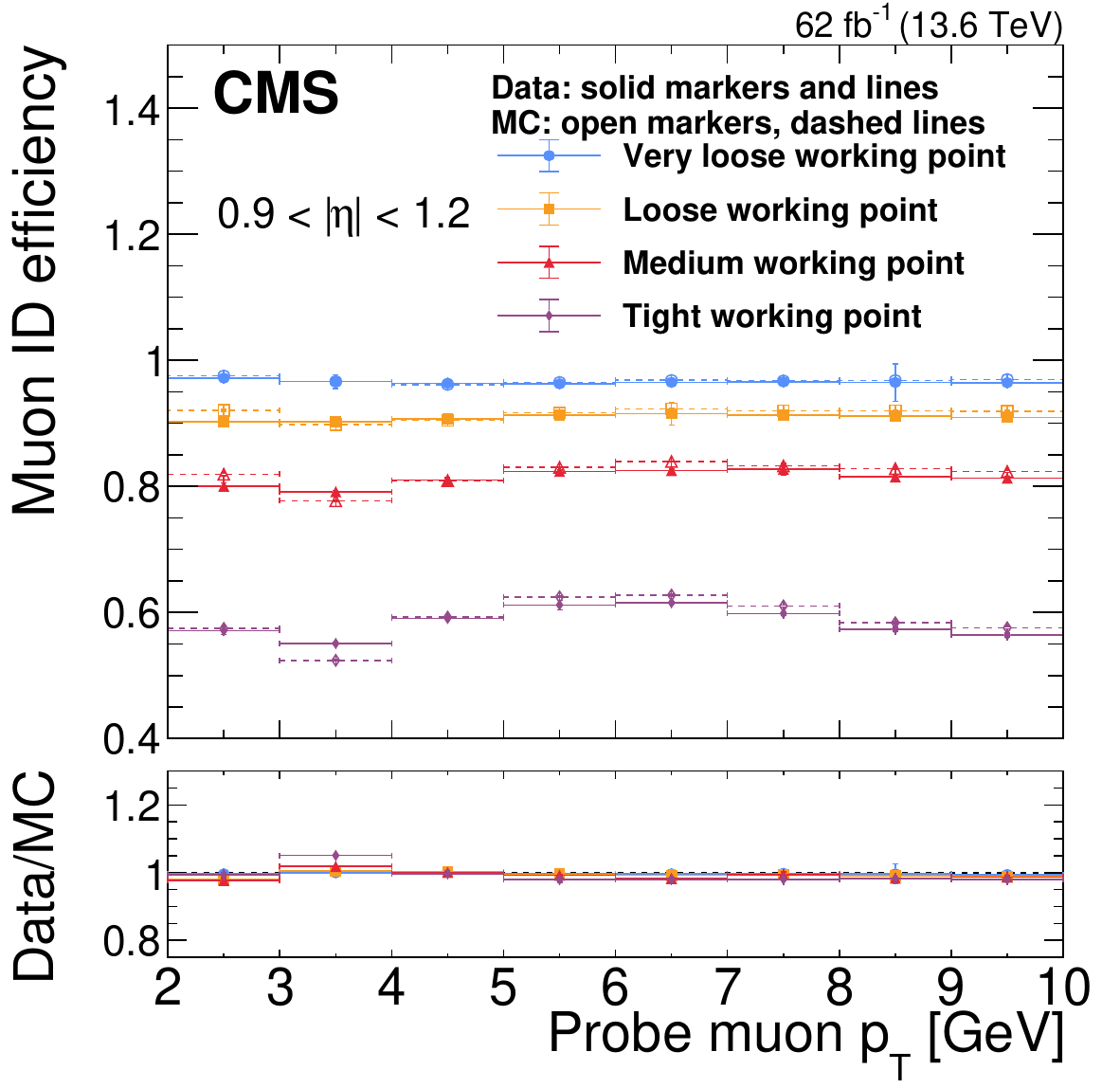}
    \includegraphics[width=0.49\textwidth]{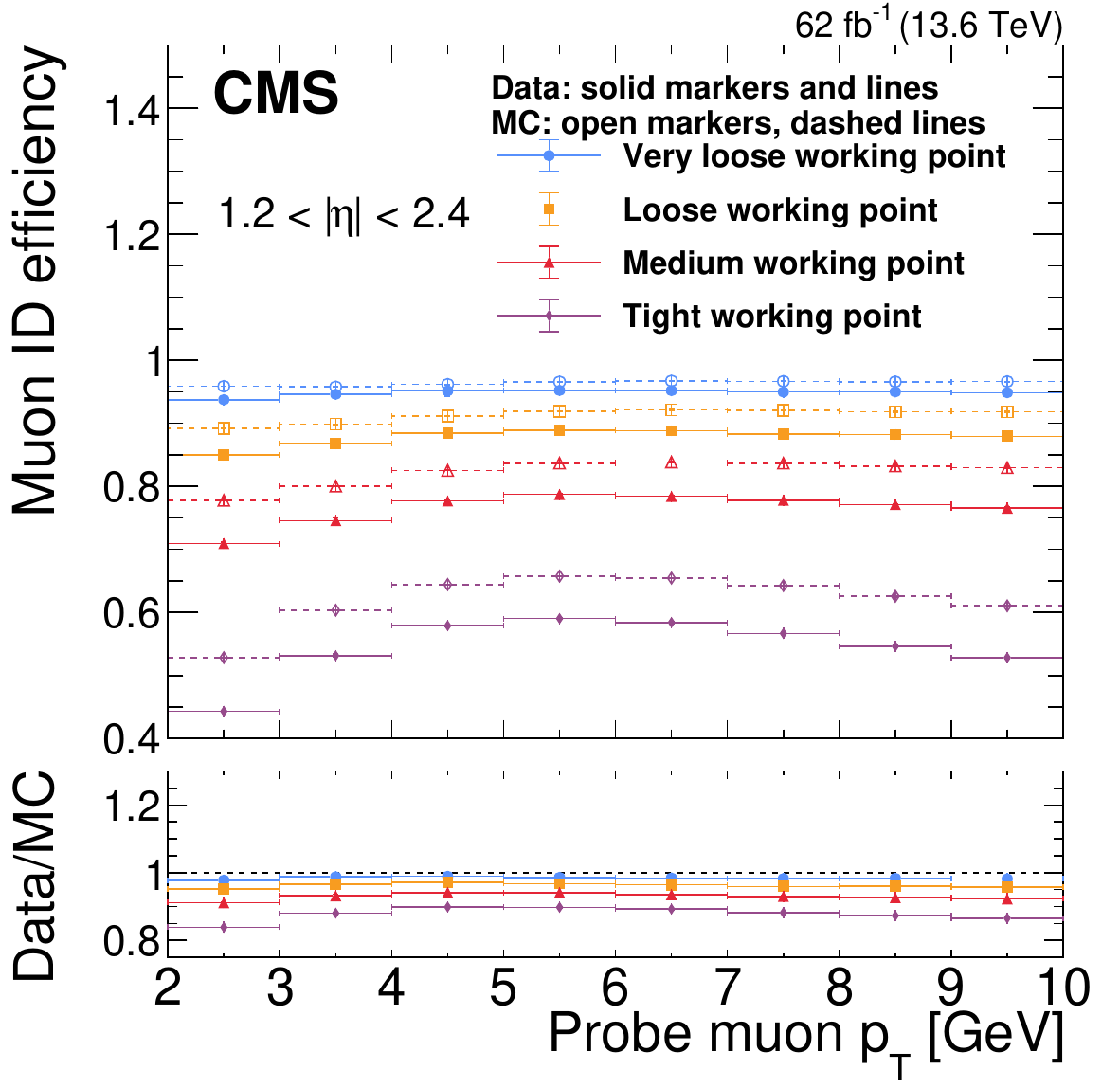}    
    \caption{Efficiencies of the different working points of the Run~3 soft-muon MVA as functions of muon \pt for muon $0 < \abseta < 0.9$ (upper left), $0.9 < \abseta < 1.2$ (upper right), $1.2 < \abseta < 2.4$ (lower). The vertical bars indicate the total uncertainty including statistical and systematic uncertainties.}
    \label{fig:effptWPs}
\end{figure}

\begin{figure}
    \centering
    \includegraphics[width=0.49\textwidth]{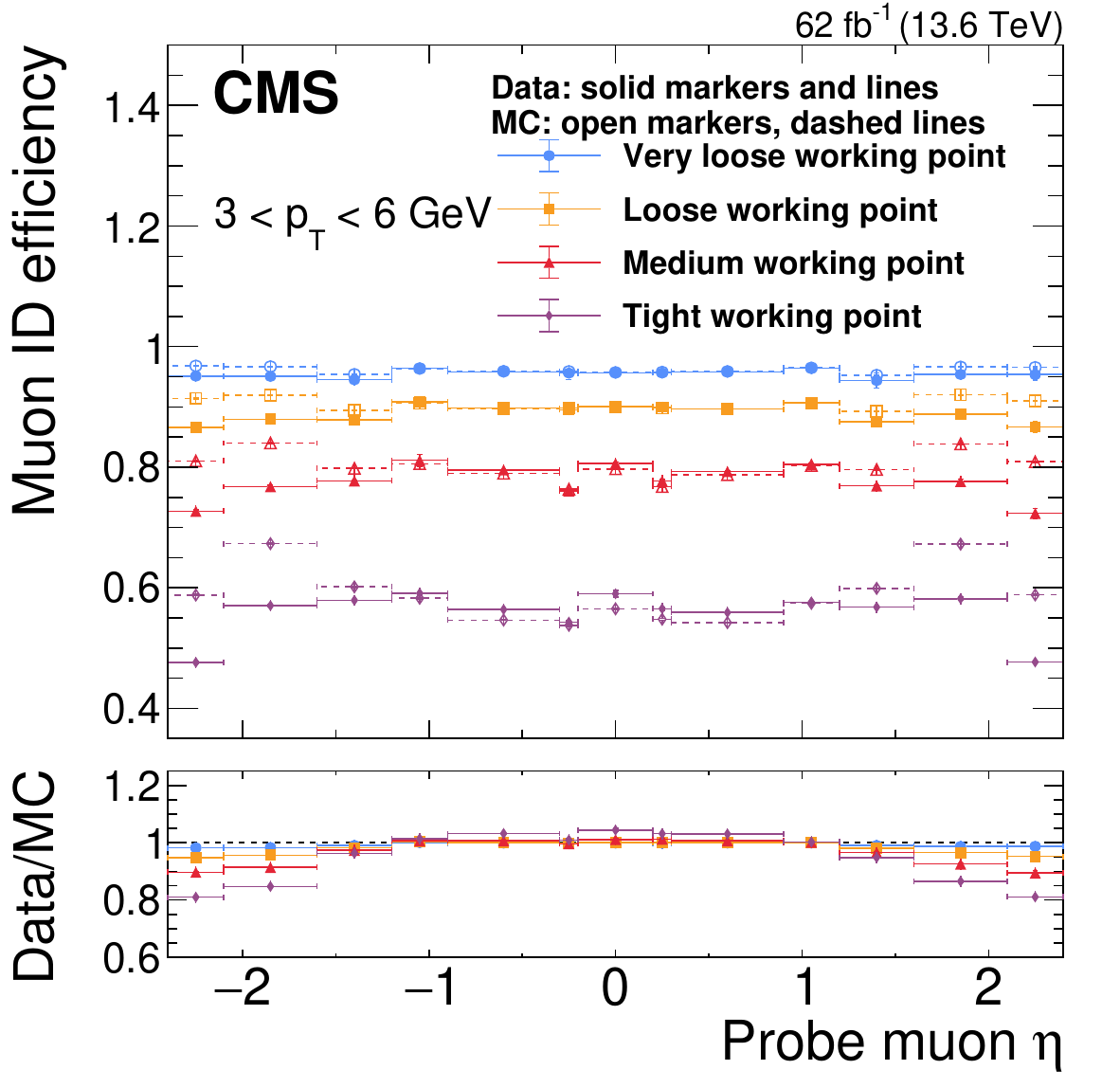}   
    \includegraphics[width=0.49\textwidth]{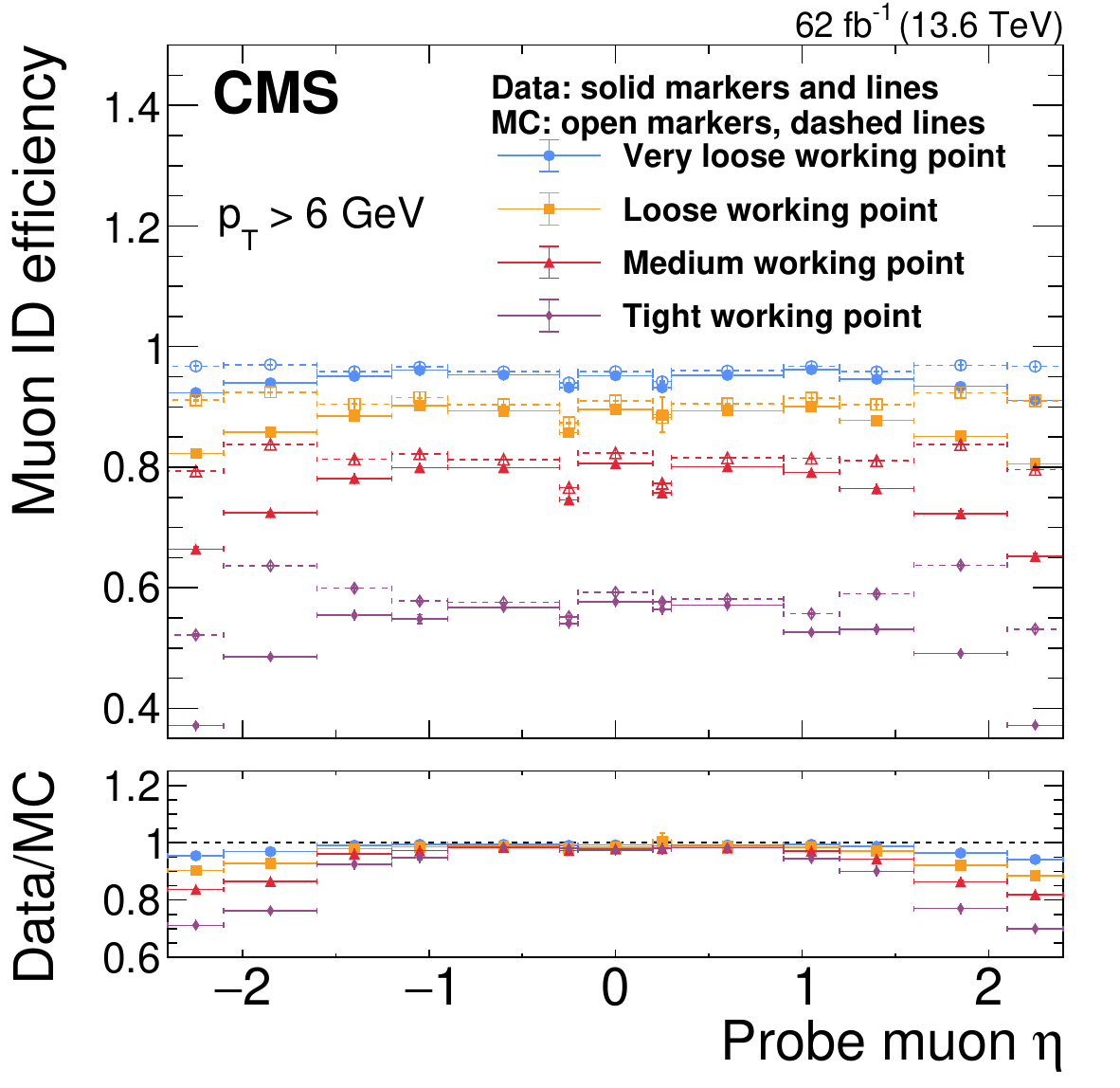}    
    \caption{Efficiencies of different working points of the Run~3 soft-muon MVA as functions of muon $\eta$ for muons with $3 < \pt < 6\GeV$ (left) and $\pt > 6\GeV$ (right). The vertical bars indicate the total uncertainty including statistical and systematic uncertainties.}
    \label{fig:effetaWPs}
\end{figure}

In Figs.~\ref{fig:effptWPs} and~\ref{fig:effetaWPs}, the efficiency is shown for the four different working points defined for the Run~3 soft-muon MVA. The very loose working point provides an efficiency of about 95\% and has very stable performance as a function of both \pt and $\eta$. The loose working point provides efficiencies between 82 and 90\%, depending on $\eta$, with the efficiency dropping off towards large values of \abseta. For the medium working point, a similar spread of efficiency values is observed, ranging from 68 to 80\%, with a more pronounced drop-off towards high \abseta. Most affected by the inefficiency in the forward region is the tight working point, for which the efficiency ranges from 42 to 60\%. For all working points, the efficiency is very well modeled in simulation in the central region of the detector and is overestimated in simulation with respect to data in the forward region because of the mismodeling of the input features in the simulation and issues with the readout of the CSC detectors, as discussed above.

\subsection{Background rate} \label{sec:bkgRate}
The rate at which background muons pass the different working points of the Run~3 soft-muon MVA is measured separately in data and simulation for muons from \PGp and \PK meson decays. 

\subsubsection{Background rate for \texorpdfstring{\pmb{\pimunu}}{pi to mu-nu} decays}

A sample of \ks meson decays into a pair of charged pions is used to evaluate the probability of a muon from a \PGp meson decay to pass the Run~3 soft-muon MVA. The events are collected using a single-electron trigger requiring $\pt > 30\GeV$. The resulting samples consist predominantly of leptonically decaying $\PW$ and $\PZ$ bosons as well as \ttbar decays where the hadronic activity in the events is a source of \ks decays unbiased by the trigger selection. The \PGp mesons are reconstructed as charged particle tracks in the CMS detector that are required to have $\pt > 4\,(1)\GeV$ for the \PGp meson with the higher (lower) \pt and $\abseta < 2.4$. The invariant mass of the \ks candidate is required to be between 0.45 and 0.55\GeV. Loose selection criteria, which do not bias the measured background rate, are applied. The \PGp tracks are fitted to a common vertex and the decay length of the \ks divided by its uncertainty is required to be larger than 3 and the probability of the two tracks to originate from the same vertex must be larger than 0.001. The $\chi^2$ divided by the degrees of freedom of the common vertex fit is required to be less than 3. The cosine of the angle between the momentum direction of the \ks candidate and the vector from the beam spot to the decay vertex in the transverse plane is required to be greater than 0.999. The three-dimensional impact parameter significance, defined as the value of the impact parameter divided by its uncertainty, of the \ks candidate with respect to the primary vertex is required to be less than 3 and the two-dimensional impact parameter significance of each pion track with respect to the beam spot is required to be greater than 5. All pion and muon tracks are required to pass the ``high purity'' flag. This is one of the input variables of the Run~3 soft-muon MVA, but the effect of including it in the selection for the background rate measurement is negligible.

\begin{figure}[htb]
    \centering
    \includegraphics[width=0.49\textwidth]{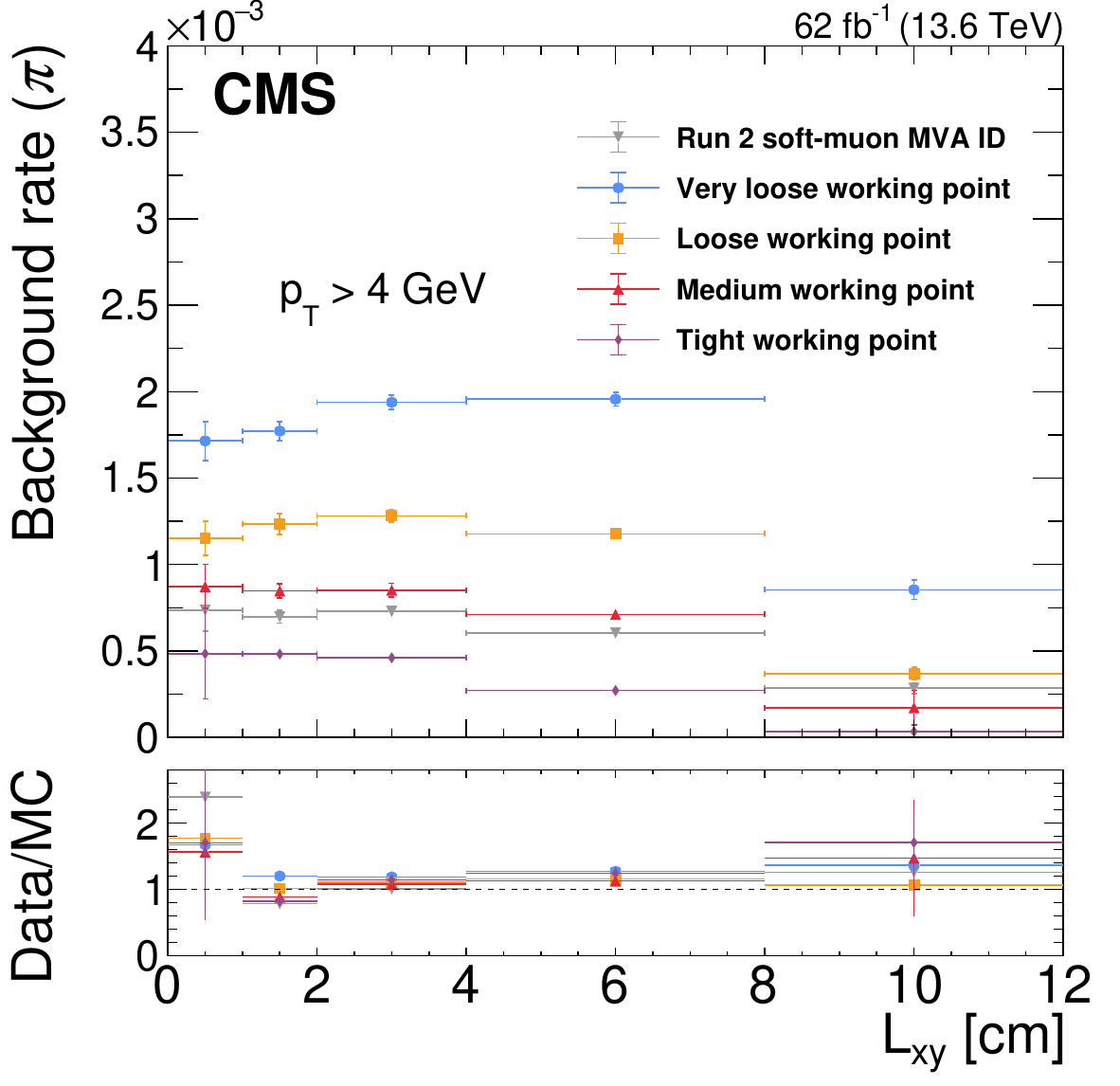}
    \includegraphics[width=0.49\textwidth]{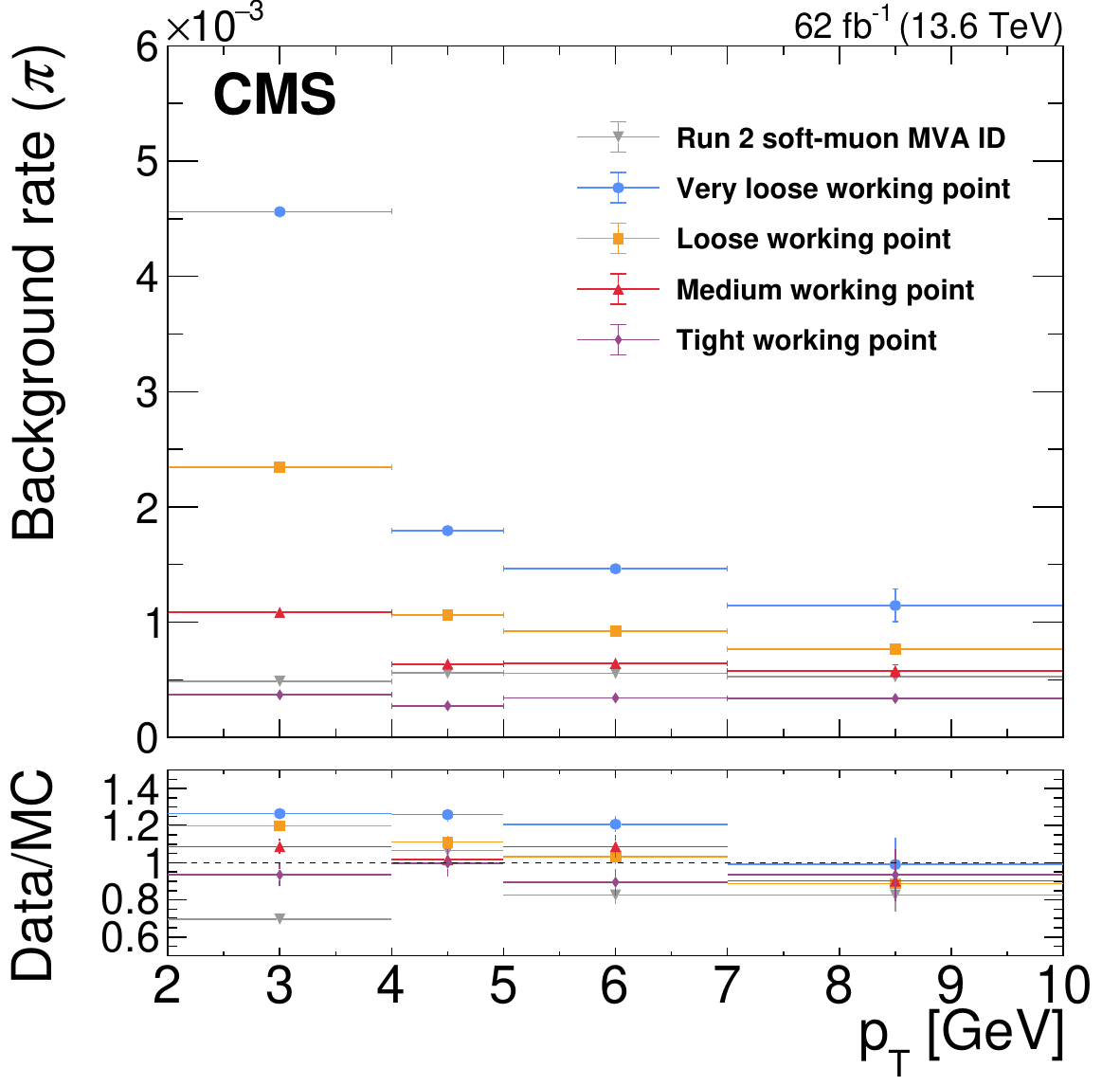}
    \caption{Background rates for muons from pions of different working points of the Run~3 soft-muon MVA for 2022--2023 data and simulation as a function of pion decay length ($L_{xy}$) measured in cm (left) and muon \pt (right). The muon $\pt>4\GeV$ is required in the $L_{xy}$ plot, whereas the misidentification rate of muons $2<\pt<4\GeV$ is measured only for $\abseta > 1$ since central muons in this \pt range do not reach the muon detector. The vertical bars indicate the statistical uncertainty.}
    \label{fig:fakerate_pion}
\end{figure}

We distinguish the cases of all \kspipi and those where one pion decays to a muon that passes a working point of the soft-muon MVA (\ksmupi). The background rate is defined as the ratio of \ksmupi decays to the total number of \kspipi decays. To extract the yield of \ks candidates in the two categories, the distribution of the invariant mass of the \ks candidate is fitted in both cases with a probability density function that models the \ks peak with a sum of Crystal Ball~\cite{Crystal} and Gaussian functions with the same mean, and it uses second-order Bernstein polynomials to describe the combinatorial background. In the \ksmupi case, the shape of the \ks peak is fixed to that fitted for \kspipi events in the bins with low event counts to ensure a stable result. The lifetime of the \ks meson is significantly larger than that of \PQb and \PQc hadrons. Therefore, the background rate is measured as a function of the kaon decay length $L_{xy}$, defined as the distance between the beam spot and the kaon decay vertex in the transverse plane. The results, together with the background rate as a function of muon \pt are shown in Fig.~\ref{fig:fakerate_pion}. As muons in the central part of the detector are not able to reach the muon detectors if their \pt is very low, $\pt > 4\GeV$ is required for the measurement as a function of $L_{xy}$, and $\abseta > 1$ is required for the measurement as a function of \pt, for $\pt < 4\GeV$. The background rate is mostly stable as a function of both quantities, especially for the medium and tight working points. As designed, the background rate for the medium working point is very similar to that for the Run~2 soft-muon MVA, the latter being slightly lower overall. An exception to this is the bin below 4\GeV, where the Run~2 soft-muon MVA shows a significantly lower background rate. This matches the much lower efficiency observed in this \pt range compared to the Run~3 version. The background rate drops for all working points for $L_{xy} > 8\cm$. For the very loose and loose working points, the background rate is somewhat higher for very low muon \pt, while it is flatter for the medium and tight working points. No dependence of the background rate on the evolving detector conditions during 2022 and 2023 is observed within the statistical uncertainties of the measurement. The same holds for the background rate for \kmunu decays described below.

\subsubsection{Background rate for \texorpdfstring{\pmb{\kmunu}}{K to mu-nu} decays}

The background rate for \PK mesons is measured similarly to that for \PGp mesons using \phikk decays. Events are again required to have passed the single-electron trigger with $\pt > 30\GeV$. As before, a set of loose selection criteria is applied. The \PK candidates are reconstructed as charged particle tracks that are required to have $\pt > 4\,(3)\GeV$ for the \PK meson with the higher (lower) \pt and $\abseta < 2.4$, and the \PGf candidate is required to fulfill $1.00 < m_{\PGf} < 1.04\GeV$. The probability of the two kaons to originate from the same vertex is required to be greater than 0.3. The three-dimensional impact parameter of the \PGf candidate with respect to the primary vertex is required to be less than 1\cm. The distance of the closest approach between the two kaon tracks is required to be less than 0.004\cm.

\begin{figure}[htb]
    \centering
    \includegraphics[width=0.49\textwidth]{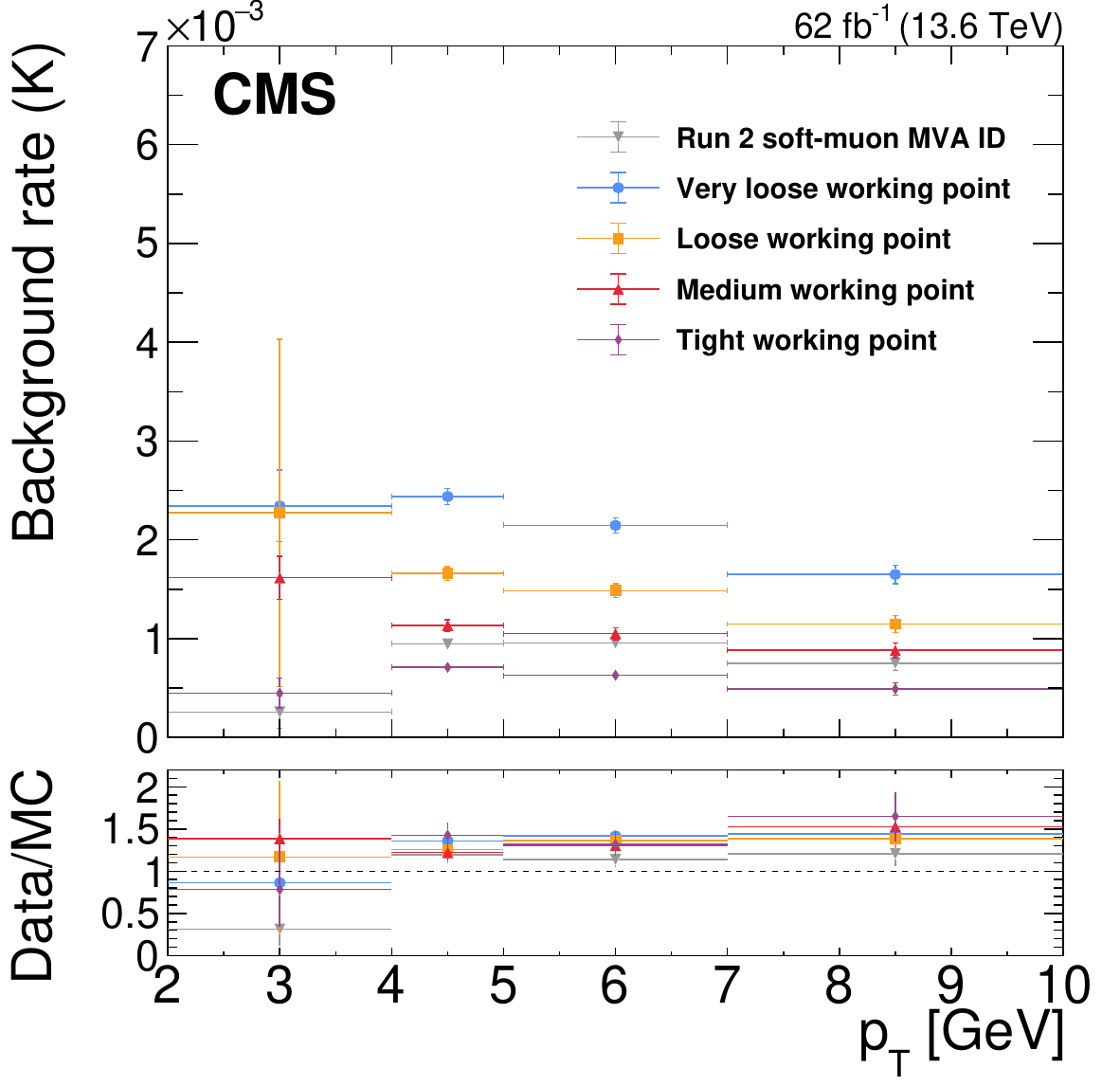}
    \caption{Background rates for muons from kaons of different working points of the Run~3 soft-muon MVA for 2022--2023 data and simulation as a function of muon \pt. The misidentification rate of muons $2<\pt<4\GeV$ is measured only for $\abseta > 1$ since central muons in this \pt range do not reach the muon detector. The vertical bars indicate the statistical uncertainty.}
    \label{fig:fakerate_kaon}

\end{figure}

A similar fit procedure as for the \PGp case is performed to extract the \phikk and \phimuk signal yields on top of a large combinatorial background. In this case, the signal peak is modeled with a Voigtian function, while the background is again modeled with a second-order Bernstein polynomial. Again, the shape of the signal peak in the \phimuk case is fixed to that obtained in the \phikk case. The kaons that decay into muons in flight tend to have much smaller decay lengths than those of pions. Therefore, the background rate is measured only as a function of muon \pt, as shown in Fig.~\ref{fig:fakerate_kaon}. As before, $\abseta > 1$ is required for $2 < \pt < 4\GeV$. In this case, all working points exhibit a downward trend for the background rate with increasing muon \pt.

\section{Summary}
\label{sec:summary}
A multivariate (MVA) classifier based on gradient-boosted decision trees has been developed for the selection of soft muons from the decay of heavy-flavor hadrons and various rare decays for transverse muon momenta \pt below 10\GeV. It is optimized for the analysis of data recorded by the CMS experiment during Run~3 of the LHC. The classifier was trained to separate these muons from background muons from pion and kaon decays. The training uses muons that pass a looser selection in a larger phase space, compared with the training of a similar classifier used in the analysis of Run~2 data, thus increasing the sensitivity to the processes outlined above. The new training also takes into account a larger set of input features and uses input samples matching the beam and detector conditions in Run~3. Consequently, the new Run~3 soft-muon MVA offers significantly improved selection efficiency for the same background rejection as the Run~2 version in both simulated samples and collision data recorded in 2022 and 2023. This improvement is most apparent for muons with $\pt < 4\GeV$ and muons reconstructed in the forward direction of the CMS detector, which were not included in the training of the Run~2 version of the classifier. The efficiency and background rate of the Run~3 soft-muon MVA measured in collision data is generally well described by the CMS simulation, with some significant differences observed for forward muons, especially for strict selections on the MVA score, which can be corrected at the analysis level.

\begin{acknowledgments}
We congratulate our colleagues in the CERN accelerator departments for the excellent performance of the LHC and thank the technical and administrative staffs at CERN and at other CMS institutes for their contributions to the success of the CMS effort. In addition, we gratefully acknowledge the computing centers and personnel of the Worldwide LHC Computing Grid and other centers for delivering so effectively the computing infrastructure essential to our analyses. Finally, we acknowledge the enduring support for the construction and operation of the LHC, the CMS detector, and the supporting computing infrastructure provided by the following funding agencies: SC (Armenia), BMBWF and FWF (Austria); FNRS and FWO (Belgium); CNPq, CAPES, FAPERJ, FAPERGS, and FAPESP (Brazil); MES and BNSF (Bulgaria); CERN; CAS, MoST, and NSFC (China); MINCIENCIAS (Colombia); MSES and CSF (Croatia); RIF (Cyprus); SENESCYT (Ecuador); ERC PRG, RVTT3 and MoER TK202 (Estonia); Academy of Finland, MEC, and HIP (Finland); CEA and CNRS/IN2P3 (France); SRNSF (Georgia); BMBF, DFG, and HGF (Germany); GSRI (Greece); NKFIH (Hungary); DAE and DST (India); IPM (Iran); SFI (Ireland); INFN (Italy); MSIP and NRF (Republic of Korea); MES (Latvia); LMTLT (Lithuania); MOE and UM (Malaysia); BUAP, CINVESTAV, CONACYT, LNS, SEP, and UASLP-FAI (Mexico); MOS (Montenegro); MBIE (New Zealand); PAEC (Pakistan); MES and NSC (Poland); FCT (Portugal); MESTD (Serbia); MCIN/AEI and PCTI (Spain); MOSTR (Sri Lanka); Swiss Funding Agencies (Switzerland); MST (Taipei); MHESI and NSTDA (Thailand); TUBITAK and TENMAK (Turkey); NASU (Ukraine); STFC (United Kingdom); DOE and NSF (USA).

\hyphenation{Rachada-pisek} Individuals have received support from the Marie-Curie program and the European Research Council and Horizon 2020 Grant, contract Nos.\ 675440, 724704, 752730, 758316, 765710, 824093, 101115353, 101002207, and COST Action CA16108 (European Union); the Leventis Foundation; the Alfred P.\ Sloan Foundation; the Alexander von Humboldt Foundation; the Science Committee, project no. 22rl-037 (Armenia); the Belgian Federal Science Policy Office; the Fonds pour la Formation \`a la Recherche dans l'Industrie et dans l'Agriculture (FRIA-Belgium); the F.R.S.-FNRS and FWO (Belgium) under the ``Excellence of Science -- EOS" -- be.h project n.\ 30820817; the Beijing Municipal Science \& Technology Commission, No. Z191100007219010 and Fundamental Research Funds for the Central Universities (China); the Ministry of Education, Youth and Sports (MEYS) of the Czech Republic; the Shota Rustaveli National Science Foundation, grant FR-22-985 (Georgia); the Deutsche Forschungsgemeinschaft (DFG), among others, under Germany's Excellence Strategy -- EXC 2121 ``Quantum Universe" -- 390833306, and under project number 400140256 - GRK2497; the Hellenic Foundation for Research and Innovation (HFRI), Project Number 2288 (Greece); the Hungarian Academy of Sciences, the New National Excellence Program - \'UNKP, the NKFIH research grants K 131991, K 133046, K 138136, K 143460, K 143477, K 146913, K 146914, K 147048, 2020-2.2.1-ED-2021-00181, TKP2021-NKTA-64, and 2021-4.1.2-NEMZ\_KI-2024-00036 (Hungary); the Council of Science and Industrial Research, India; ICSC -- National Research Center for High Performance Computing, Big Data and Quantum Computing and FAIR -- Future Artificial Intelligence Research, funded by the NextGenerationEU program (Italy); the Latvian Council of Science; the Ministry of Education and Science, project no. 2022/WK/14, and the National Science Center, contracts Opus 2021/41/B/ST2/01369 and 2021/43/B/ST2/01552 (Poland); the Funda\c{c}\~ao para a Ci\^encia e a Tecnologia, grant CEECIND/01334/2018 (Portugal); the National Priorities Research Program by Qatar National Research Fund; MCIN/AEI/10.13039/501100011033, ERDF ``a way of making Europe", and the Programa Estatal de Fomento de la Investigaci{\'o}n Cient{\'i}fica y T{\'e}cnica de Excelencia Mar\'{\i}a de Maeztu, grant MDM-2017-0765 and Programa Severo Ochoa del Principado de Asturias (Spain); the Chulalongkorn Academic into Its 2nd Century Project Advancement Project, and the National Science, Research and Innovation Fund via the Program Management Unit for Human Resources \& Institutional Development, Research and Innovation, grant B39G670016 (Thailand); the Kavli Foundation; the Nvidia Corporation; the SuperMicro Corporation; the Welch Foundation, contract C-1845; and the Weston Havens Foundation (USA).
\end{acknowledgments}

\bibliography{auto_generated}
\cleardoublepage \appendix\section{The CMS Collaboration \label{app:collab}}\begin{sloppypar}\hyphenpenalty=5000\widowpenalty=500\clubpenalty=5000
\cmsinstitute{Yerevan Physics Institute, Yerevan, Armenia}
{\tolerance=6000
V.~Chekhovsky, A.~Hayrapetyan, V.~Makarenko\cmsorcid{0000-0002-8406-8605}, A.~Tumasyan\cmsAuthorMark{1}\cmsorcid{0009-0000-0684-6742}
\par}
\cmsinstitute{Institut f\"{u}r Hochenergiephysik, Vienna, Austria}
{\tolerance=6000
W.~Adam\cmsorcid{0000-0001-9099-4341}, J.W.~Andrejkovic, L.~Benato\cmsorcid{0000-0001-5135-7489}, T.~Bergauer\cmsorcid{0000-0002-5786-0293}, S.~Chatterjee\cmsorcid{0000-0003-2660-0349}, K.~Damanakis\cmsorcid{0000-0001-5389-2872}, M.~Dragicevic\cmsorcid{0000-0003-1967-6783}, P.S.~Hussain\cmsorcid{0000-0002-4825-5278}, M.~Jeitler\cmsAuthorMark{2}\cmsorcid{0000-0002-5141-9560}, N.~Krammer\cmsorcid{0000-0002-0548-0985}, A.~Li\cmsorcid{0000-0002-4547-116X}, D.~Liko\cmsorcid{0000-0002-3380-473X}, I.~Mikulec\cmsorcid{0000-0003-0385-2746}, J.~Schieck\cmsAuthorMark{2}\cmsorcid{0000-0002-1058-8093}, R.~Sch\"{o}fbeck\cmsAuthorMark{2}\cmsorcid{0000-0002-2332-8784}, D.~Schwarz\cmsorcid{0000-0002-3821-7331}, M.~Sonawane\cmsorcid{0000-0003-0510-7010}, W.~Waltenberger\cmsorcid{0000-0002-6215-7228}, C.-E.~Wulz\cmsAuthorMark{2}\cmsorcid{0000-0001-9226-5812}
% ...
\par}
\vskip\cmsinstskip
\dag:~Deceased\\
$^{1}$Also at Yerevan State University, Yerevan, Armenia\\
$^{2}$Also at TU Wien, Vienna, Austria\\
% ...
\end{sloppypar}
\end{document}